\documentclass[aps,pre,superscriptaddress,twocolumn,longbibliography]{revtex4}
\usepackage{graphicx,epstopdf,url}
\usepackage[colorlinks=true,urlcolor=blue,citecolor=blue,linkcolor=blue,urlcolor=blue]{hyperref}
\usepackage[active]{srcltx}

\begin{document}

\title{
Multistability of graphene nanobubbles
}

\author{Alexander V. Savin}
\email{asavin@chph.ras.ru}
\affiliation{
N.N. Semenov Federal Research Center for Chemical Physics of the Russian Academy of Sciences,
4 Kosygin St., Moscow 119991, Russia}
\affiliation{
Plekhanov Russian University of Economics, 36 Stremyanny Lane, Moscow 117997, Russia
}

\begin{abstract}
Using He, Ne, Ar, Kr, and Xe atoms as a model system, it is demonstrated that graphene nanobubbles on flat substrates are multistable systems.
A nanobubble can adopt multiple stable stationary states, each characterized by the number of layers $l$ within the cluster of encapsulated atoms.
The layers are circular, concentrically stacked, and form an $l$-stepped pyramid with a flat top.
Encapsulation of this pyramid by the graphene sheet is achieved through local stretching of the membrane: the valence bonds elongate only directly above the confined atoms.
Outside this coverage zone, the sheet remains undeformed and lies flush against the substrate.
The maximum number of possible layers, $l_m$, increases monotonically with the number of encapsulated atoms $N$, reaching $l_m=6$ for $N=4000$.
The graphene membrane, through van der Waals interaction with the substrate, compresses the internal atomic cluster, generating pressures on the order of $P\sim 1$~GPa.
Numerical simulations of thermal vibrations reveal that among all $l$-layer configurations, one ground state always exist.
Upon heating, this state smoothly transitions into a layerless liquid configuration.
All other stationary states transform into this ground state once a characteristic temperature $T_l$ is reached. For $N=4000$, the ground state corresponds to the four-layer packing ($l=4$).
The coexistence of multiple stable states with distinct layer numbers at low temperatures leads to the absence of a universal shape for the nanobubbles.
In this scenario, the height-to-radius ratio, $H/R$ is not constant and can vary from 0 to 0.28, depending on the number of layers. 
\end{abstract}
\maketitle

\section{Introduction \label{sec1}}
Carbon atoms can form numerous structures, among which the monoatomic crystalline layer graphene has recently attracted significant research attention \cite{Novoselov2004,Geim2007,Soldano2010,Baimova2014a,Baimova2014b}.
This nanomaterial is of great interest due to its unique electronic \cite{Geim2009}, mechanical \cite{Lee2008}, and thermal properties \cite{Baladin2008,Liu2015}.
Another important physical and chemical property of graphene is its ability to trap gas molecules under high pressure when placed on various flat substrates.
This can lead to the formation of bubbles ranging in size from nanometers to micrometers.

High interest is also directed towards other graphene-like materials, such as sheets of hexagonal boron nitride (h-BN), fluorographene (CF), molybdenum disulfide (MoS$_2$), and tungsten selenide (WSe$_2$).
These two-dimensional materials offer vast possibilities for creating heterostructures in the form of van der Waals stacks, which possess novel properties \cite{Geim2013,Wang2013,Xiang2020,Zhang2020}.
An important feature of such heterostructures is their ability to trap molecules in localized regions between adjacent layers.
Due to van der Waals interactions between the layers, internal pressure on the order of several GPa can arise within these traps (nanopockets) \cite{Vasu2016,Khestanova2016,Hu2023}, which can significantly change the properties of the captured material.
Covering a molecular system on a flat substrate with a graphene sheet (van der Waals encapsulation) is a convenient method for creating high local pressure to modify the system's properties \cite{Zhang2017} or for its storage \cite{Zheng2012,Slepchenkov2018,Apkadirova2022}.
In graphene nanobubbles, high pressure is induced by the forces of interaction between the sheet and the substrate.

Graphene nanobubbles have the potential to become indispensable tools in nanotechnology.
Analysis of their shape allows for the determination of the adhesion energy between a graphene sheet and a substrate \cite{Wang2016}.
Nanobubbles in multilayer graphene can be utilized in the development of cathode materials for lithium-ion batteries \cite{He2023}. Furthermore, they can be used to create two-dimensional van der Waals solids, which may find applications in quantum information technologies \cite{Langle2024}.

Encapsulation structures (nanobubbles) can form spontaneously as surface defects during the mechanical transfer of two-dimensional sheets in the creation of van der Waals heterostructures \cite{Khestanova2016,Sanchez2018}.
Nanobubbles can also be intentionally fabricated using ion bombardment \cite{Blundo2020,Zamborlini2015,Villarreal2021} or selective adsorption methods \cite{Zahra2020}.

In addition to experimental techniques, analytical models and numerical simulations have become essential tools for investigating nanobubbles. Analytical approaches are often based on elastic plate theory, which incorporates the contribution of adhesion energy between the layers of van der Waals heterostructures \cite{Yue2012, Wang2013a, Dai2018}. More advanced models further integrate the equation of state of the encapsulated substance \cite{Zhilyaev2019, Iakovlev2019, Aslymov2020, Qu2023} and account for nonlinear effects \cite{Lyublinskaya2020}. These models have revealed that circular nanobubbles exhibit universal scaling behavior, characterized by a constant height-to-radius aspect ratio, $H/R$.

Numerical simulations serve not only to validate analytical predictions but also to provide deeper insights into the physical properties of nanobubbles. For instance, it has been shown that spatial confinement and high internal pressure can induce the formation of internal crystalline structures, which may exhibit planar geometries \cite{Iakovlev2017}. Furthermore, the equilibrium shape of a nanobubble can critically depend on the nature of the encapsulated material \cite{Ghorbanfekr2017}. Notably, simulations by S.K.~Jain et al. \cite{Jain2017} demonstrated that a nanobubble containing an ideal gas invariably assumes a circular shape with a constant aspect ratio of $H/R=0.204$.

The capabilities of molecular dynamics (MD) simulations have enabled the exploration of more complex phase behaviors, such as the transition of confined argon into a liquid-crystalline state \cite{Korneva2023}. MD techniques have also been employed to model the mechanical indentation of a nanobubble by an atomic force microscope probe \cite{Faraji2022}.

From a thermal management perspective, it has been demonstrated that the presence of nanobubbles can significantly reduce the interfacial thermal conductivity of layered van der Waals structures \cite{Qu2023a}. Additionally, they have been identified as a potential cause of substantial thermal contraction in these materials \cite{Qu2023b}.

In this paper, all possible stationary states of graphene nanobubbles are numerically modeled.
The paper is organized as follows.
Section \ref{sec2} describes the model employed.
Section \ref{sec3} presents the possible stationary states of the nanobubbles.
The stability of these stationary states against thermal fluctuations is analyzed in Section \ref{sec4}.
The main results are summarized in the Conclusion (Section \ref{sec5}).

\section{Model \label{sec2}}
Let us consider a two-component molecular system consisting of a graphene sheet lying on a flat substrate and atoms of inert gas placed between them.
Let the sheet comprise $N_c$ carbon atoms, and the gas component---$N_g $ argon (helium, neon, krypton, xenon) atoms.
As the substrate, we will use an attractive plane at $z = 0$, corresponding to the flat surface of a silicon oxide crystal.
This molecular system, consisting of $N=N_c+N_g$ atoms, represents the simplest model of van der Waals encapsulation.

\begin{figure}[tb]
\begin{center}
\includegraphics[angle=0, width=1.0\linewidth]{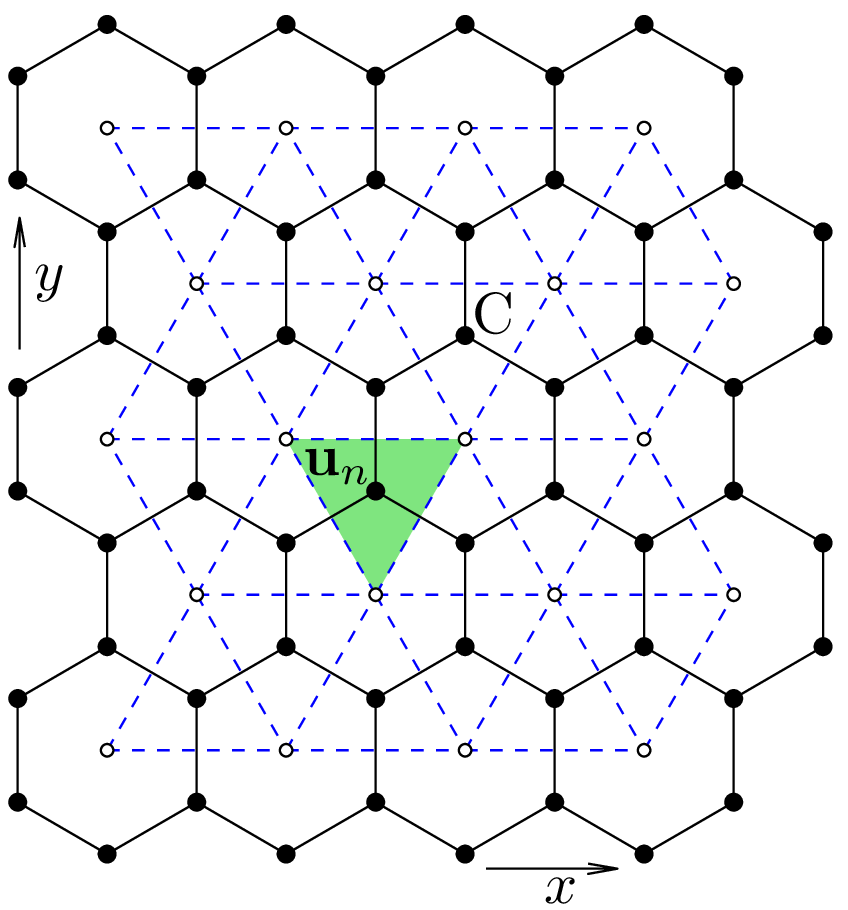}
\end{center}
\caption{\label{fg01}\protect
A rectangular graphene sheet with $N_x= 10$, $N_y=12$.
Black disks represent carbon atoms C, black solid lines represent valence bonds C--C.
Small light disks indicate the positions of the centers of mass of the valence bond hexagons.
Dashed lines illustrate the triangulation of the graphene sheet using these centers -- the inner surface of the sheet is divided into triangles, each containing one carbon atom (the triangle corresponding to the $n$-th atom is highlighted; $\mathbf{u}_n$ is the coordinate vector of the atom).
}
\end{figure}

Consider a rectangular graphene sheet in which the chains of valent bonds along the $x$-axis have a "zigzag" structure with $N_x$ nodes, and along the $y$-axis---an "armchair" structure with $N_y$ nodes (the sheet lies parallel to the $z=0$ plane).
Consider a sheet with periodic and free boundary conditions (Fig.~\ref{fg01}).
When periodic boundary conditions are applied, the sheet consists of $N_c = N_xN_y/2$ carbon atoms and has a rectangular shape of size $L_x\times L_y$, where its periodic lengths are $L_x = N_x r_c \sqrt{3} / 2$ and $L_y = 3 N_y r_c / 4$ ($r_c=1.418$~\AA{} is the C--C bond length).
When free boundary conditions are applied, the sheet consists of $N_c=N_xN_y/2-2$ carbon atoms and has a rectangular shape of size $L_x \times L_y$, where its length is $L_x=(N_x - 1)r_c\sqrt{3}/2$ and its width is $L_y=(3N_y/4-1)r_c$.
In all calculations, we use a graphene sheet with parameters $N_x=416$, $N_y=480$ (sheet dimensions: $51.086\times 51.048$~nm$^2$, number of carbon atoms $N_c = 99840$).
For the graphene sheet with free boundaries, we assume that hydrogen atoms are attached to the edge carbon atoms, which participate in the formation of only two C--C bonds.
These edge atoms, together with the attached hydrogen atoms, form united CH atoms with a mass of $M_e = 13m_p$ (the mass of the interior sheet atoms is $M_c=12m_p$, where $m_p=1.66 \times 10^{-27}$~kg is the proton mass).

The shape of the nanobubble and its properties do not depend on the boundary conditions employed.
We only note that the approach of a nanobubble to a free edge of the sheet can lead to its collapse (i.e., the release of gas atoms from under the sheet), an effect which is impossible under periodic boundary conditions.

The Hamiltonian of the nanosheet is
\begin{equation}
{\cal H}_1=\sum_{n=1}^{N_c}\left[\frac12M_c(\dot{\bf u}_n,\dot{\bf u}_n)+Q_n+Z({\bf u}_n)\right],
\label{f1}
\end{equation}
where the vector $\mathbf{u}_n=(x_n(t),y_n(t),z_n(t))$ defines the coordinates of the $n$-th carbon atom at time $t$.
The first term of the sum in equation (\ref{f1}) represents the kinetic energy, while the second term $Q_n$ describes the valence interaction of the atom with index $n$ with other atoms in the graphene sheet.
This includes deformations of valence bonds, valence angles, and torsion angles---a detailed description of the force field used is provided in \cite{Savin2010}.

The valence bond between two neighboring carbon atoms $n$ and $k$ can be described by the Morse potential
\begin{equation}
U_1({\bf u}_n,{\bf u}_k)=\epsilon_c\left\{\exp [-\alpha(r_{nk}-r_c)]-1\right\}^2,
\label{f2}
\end{equation}
where $r_{nk}= |\mathbf{u}_n - \mathbf{u}_k|$ is the distance between the atoms, and the bond energy is $\epsilon_c=4.9632$~eV.

Valence angle deformation energy between three adjacent carbon atoms $n$, $k$, and $l$ can be described by the potential
\begin{equation}
U_2({\bf u}_n,{\bf u}_k,{\bf u}_l)=\epsilon_\varphi(\cos\varphi-\cos\varphi_0)^2,
\label{f3}
\end{equation}
where $\cos\varphi=({\bf u}_n-{\bf u}_k,{\bf u}_l-{\bf u}_k)/r_{nk}r_{kl}$,
and $\varphi_0=2\pi/3$ is the equilibrium valent angle.
Parameters $\alpha=1.7889$~\AA$^{-1}$ and $\epsilon_\varphi=1.3143$~eV can be found
from the small amplitude oscillations spectrum of the graphene sheet \cite{Savin2008}.

Valence bonds between four adjacent carbon atoms $n$, $m$, $k$ , and $l$ constitute torsion angles, the potential energy of which can be defined as
\begin{equation}
U_3({\bf u}_n,{\bf u}_m,{\bf u}_k,{\bf u}_l)=\epsilon_\phi(1-\cos\phi),
\label{f4}
\end{equation}
where $\phi$ is the corresponding torsion angle ($\phi=0$ is the equilibrium value of the angle)
and $\epsilon_\phi=0.499$~eV.

The last term in the sum (\ref{f1}) describes the interaction of the sheet atoms with the flat substrate on which it rests.
Let the flat substrate occupy the half-space $z \le 0$.
In this case, the interaction energy of a carbon atom with the substrate can be described using a (3,9) Lennard-Jones potential \cite{Aitken2010,Zhang2013,Zhang2014}:
\begin{equation}
Z({\bf u}) = Z(z) = \varepsilon_0 \left[ (h_0 / z)^9 - 3 (h_0 / z)^3 \right] / 2,
\label{f5}
\end{equation}
where $\varepsilon_0$ is the interaction energy (adhesion energy) and $h_0$ is the equilibrium distance to the surface of the half-space.
For carbon atom and silicon dioxide (SiO$_2$) substrate, the energy is $\varepsilon_0 = 0.074$~eV and the distance is $h_0= 5$~\AA~ \cite{Koenig2011}.
The values used for the remaining atoms are presented in Table \ref{tab1}; the parameters were obtained using the Lorentz-Berthelot mixing rules.
\begin{table}[tb]
\caption{
Values of the parameters of the interaction potential with a flat substrate (\ref{f5}) and masses for atoms C, He, Ne, Ar, Kr, Xe.
\label{tab1}
}
\begin{center}
\begin{tabular}{c|cccccc}
\hline\hline
 ~                     & C     & He      &  Ne     & Ar    & Kr      & Xe  \\
\hline
 $\varepsilon_0$~(eV)~ &~0.074~&~0.04322~&~0.0797~&~0.1406~&~0.17242~&~0.19462~\\
 $h_0$~(\AA)           & 5.0   & 4.577   & 4.662  & 5.0    & 5.116   & 5.397 \\
 $M_g$~($m_p$)           & 12.0  & 4.0     & 20.18  & 39.95  & 83.80   & 131.29\\
\hline\hline
\end{tabular}
\end{center}
\end{table}

The Hamiltonian of the inert gas molecular subsystem has the form
\begin{equation}
{\cal H}_2=\sum_{n=N_c+1}^{N}\left[\frac12 M_g(\dot{\bf u}_n,\dot{\bf u}_n)+Z({\bf u}_n)\right]+E_2,
\label{f6}
\end{equation}
where $N= N_c + N_g$ is the total number of atoms in the two-component molecular system, the vectors $\{{\bf u}_n\}_{n=N_c+1}^N$ specify the coordinates of the gas atoms, and $M_g$ is the atomic mass (see Table \ref{tab1}).
The first term in the sum (\ref{f6}) represents the kinetic energy of the gas subsystem, the second term---the interaction energy of the gas atoms with the flat substrate.
The last term,
\begin{equation}
E_2=\sum_{n=N_c+1}^{N-1}\sum_{k=n+1}^N V(r_{nk}),
\label{f7}
\end{equation}
defines the interaction energy between the gas atoms, where $r_{nk}=|{\bf u}_n-{\bf u}_k|$ is the distance between atoms $n$ and $k$.

The interaction between atoms is described by the Lennard-Jones (12-6) potential:
\begin{equation}
V(r)=4\epsilon[(\sigma/r)^{12}-(\sigma/r)^6].
\label{f8}
\end{equation}
The values of the interaction potential parameters for various pairs of atoms are given in Table \ref{tab2}.
The LJ interactions were truncated at a cutoff distance of 30~\AA{}.
\begin{table}[tb]
\caption{
Values of the interaction potential parameters (\ref{f8}) for various pairs of atoms  \cite{Iakovlev2017,Rappe1992,Vogt2014}. \label{tab2}
}
\begin{center}
\begin{tabular}{c|ccccc}
\hline\hline
 ~                     & CHe     & CNe     & CAr      & CKr     & CXe\\
\hline
 $\varepsilon_1$~(eV)~ &~0.00161~&~0.00297~& ~0.00525~&~0.00643~&~0.00724~\\
 $\sigma$~(\AA)        & 3.0165  & 3.0915  &  3.3965  & 3.4965  & 3.7465  \\
 \hline
 ~                     & HeHe    & NeNe   & ArAr  & KrKr  & XeXe \\
\hline
 $\varepsilon_1$~(eV)~ & 0.00094 & 0.0032 & 0.010 & 0.015 & 0.019~\\
 $\sigma$~(\AA)        & 2.64    & 2.79   & 3.40  & 3.60  & 4.10 \\
\hline\hline
\end{tabular}
\end{center}
\end{table}

The interaction energy between the gas atoms and the graphene sheet is given by the sum
\begin{equation}
E_3 = \sum_{n=1}^{N_c} \sum_{k=N_c+1}^N V(r_{nk}).
\label{f9}
\end{equation}

The total Hamiltonian of the two-component molecular system is
\begin{equation}
\mathcal{H} = \mathcal{H}_1 + \mathcal{H}_2 + E_3.
\label{f10}
\end{equation}
The potential energy of the molecular system is
\begin{equation}
E = \sum_{n=1}^{N_c} Q_n + \sum_{n=1}^N Z(\mathbf{u}_n) + E_2 + E_3.
\label{f11}
\end{equation}
\begin{figure}[tb]
\begin{center}
\includegraphics[angle=0, width=0.90\linewidth]{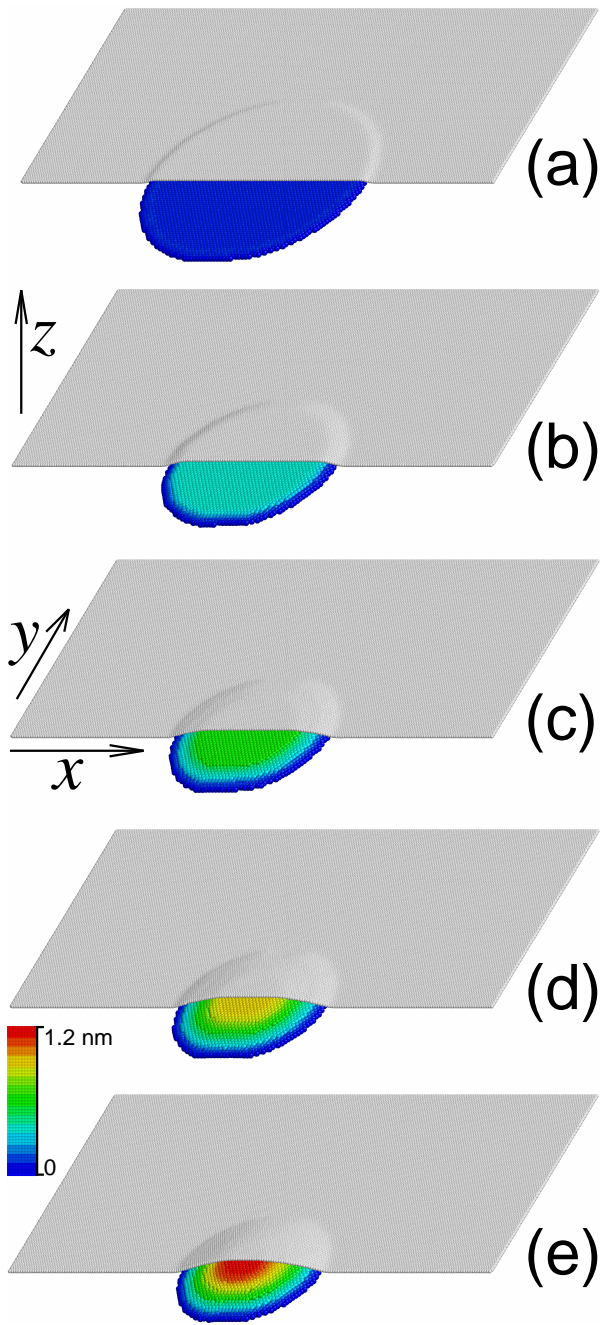}
\end{center}
\caption{\label{fg02}\protect
Stationary $l$-layer states of encapsulated $N_g=4000$ argon atoms with $l = 1$ (a), 2 (b), 3 (c), 4 (d), and 5 (e).
For clarity of presentation, only one half of the graphene sheet is shown.
The graphene sheet is depicted in gray.
For argon atoms, their vertical displacement is indicated by color.
}
\end{figure}
\begin{figure}[tb]
\begin{center}
\includegraphics[angle=0, width=1.0\linewidth]{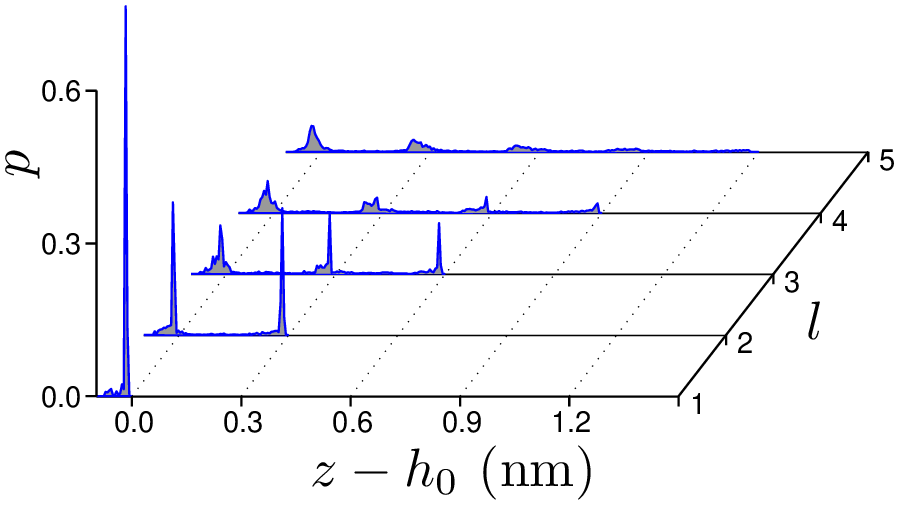}
\end{center}
\caption{\label{fg03}\protect
Distribution of vertical displacements $p(z)$ for the stationary states of encapsulated $N_g=4000$ argon atoms with number of layers $l = 1$, ..., 5.
}
\end{figure}

More detailed discussion and motivation of our choice of the interaction potentials (\ref{f2}), (\ref{f3}), (\ref{f4}) can be found in \cite{Savin2010}.
Note that in molecular dynamics of carbon nanotubes and nanoribbons, more complex reactive potentials are often used, such as Tersoff \cite{Lindsay2010}, REBO \cite{Stuart2000,Brener2002} and ReaxFF \cite{Srinivasan2015}. These potentials are able to describe the formation and breaking of bonds between carbon atoms,
but they are worse in describing the frequency spectrum of graphene and are more complicated for numerical modeling.
The use of these potentials is justified only when modeling processes in graphene-based systems that occur with changes in the topology of the valence bond lattice.
Nevertheless, these potentials reliably enough describe elastic properties of graphene \cite{Lebedeva2019}, so the decision on use them will lead to the same results.

\section{Stationary States of Graphene Nanobubbles \label{sec3}}
To find a stationary state of an encapsulated molecular system (nanobubble), it is necessary to numerically solve the potential energy minimization problem
\begin{equation}
E \rightarrow \min: \{ {\bf u}_n\}_{n=1}^N.
\label{f12}
\end{equation}
The problem (\ref{f12}) was solved numerically using the conjugate gradient method \cite{Fletcher1964,Shanno1976}.

As initial configurations, structures in the form of an $l$-layer stepped pyramid of gas atoms ($l = 1, 2, \ldots$) covered from above by a graphene sheet were used.
By varying the initial configuration, all possible stable stationary states of the encapsulated molecular system can be obtained.

Let $\{ {\bf u}_n^0 = (x_n^0, y_n^0, z_n^0)\}_{n=1}^N$ denote a solution to problem (\ref{f12}).
The corresponding stationary state is characterized by its total energy $E(\{{\bf u}_n^0\}_{n=1}^N)$, the maximum vertical displacement of the graphene sheet above the encapsulated group of atoms (i.e., the height of the graphene bulge)
$$
H = \max_{n=1,\ldots,N_c} (z_n^0 - h_0),
$$
the radius $R$ of the bulge boundary, as well as the density $d$ and pressure $P$ inside the nanobubble.
\begin{figure}[tb]
\begin{center}
\includegraphics[angle=0, width=0.73\linewidth]{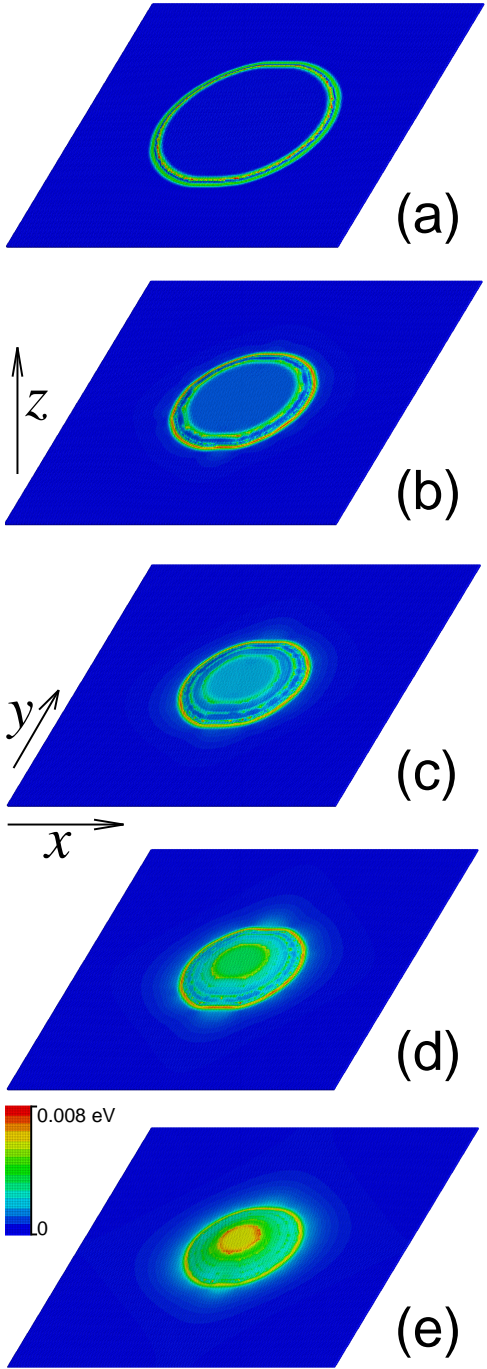}
\end{center}
\caption{\label{fg04}\protect
Stationary $l$-layer states of a nanobubble with $N_g = 4000$ krypton atoms at $l = 1$ (a), 2 (b), 3 (c), 4 (d), and 5 (e).
A top view is shown.
Color indicates the distribution of deformation energy across the graphene sheet.
}
\end{figure}
\begin{figure}[tb]
\begin{center}
\includegraphics[angle=0, width=0.725\linewidth]{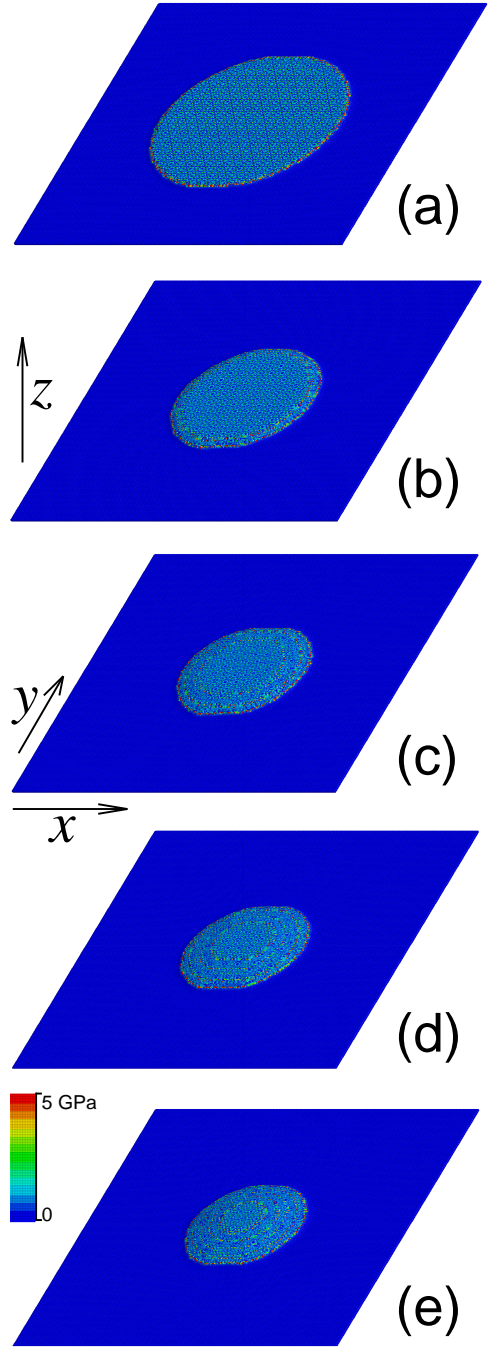}
\end{center}
\caption{\label{fg05}\protect
Stationary $l$-layer states of a nanobubble with $N_g = 4000$ xenon atoms at $l = 1$ (a), 2 (b), 3 (c), 4 (d), and 5 (e).
A top view is shown.
Color indicates the distribution of pressure across the graphene sheet.
The average pressure values are $P = 0.976$, 0.948, 0.977, 0.989,  and 0.995~GPa.
}
\end{figure}

We consider a sheet atom to participate in forming the upper surface of the nanobubble if its vertical displacement $\Delta z_n = z_n^0 - h_0 > h_1$, where the displacement threshold $h_1$ is 0.97, 1.17, 1.5, 1.6, and 1.8~\AA~ for He, Ne, Ar, Kr, and Xe atoms, respectively.
The center of the nanobubble then has coordinates
$$
x_c = \frac{1}{N_i} \sum_{n=1\atop \Delta z_n > h_1}^{N_c} x_n^0, \quad
y_c = \frac{1}{N_i} \sum_{n=1\atop \Delta z_n > h_1}^{N_c} y_n^0,
$$
where the number of graphene sheet atoms forming the upper surface of the bubble is
$$
N_i = \sum_{n=1\atop \Delta z_n > h_1}^{N_c} 1.
$$
\begin{table}[tb]
\caption{
Dependence on the number of encapsulated helium atoms $N_g$ and the number of layers $l$ of the relative specific energy $\Delta E$, pressure $P$, height $H$, radius $R$, and density $d$ of the stationary state of the nanobubble.
The number of layers in the ground state is highlighted in bold.
\label{tab3}
}
\begin{center}
\begin{tabular}{c|ccccccc}
\hline\hline
$N_g$& ~~$l$~&$\Delta E$~(eV)& $P$~(GPa) & $H$~(\AA) &  $R$~(\AA) &  $d$~(g/cm$^3$) & $H/R$\\
\hline
125  &{\bf 1} & 0      & 1.933 & 2.19 &  13.58  & 0.52 & 0.161\\
\hline
250  &{\bf 1} &   0    & 1.641 & 2.12 &  19.86  & 0.55 & 0.107\\
~    & 2      & 0.0076 & 1.474 & 3.82 &  19.09  & 0.50 & 0.200\\
\hline
500  & 1      & 0      & 1.422 & 2.11 &  28.31  & 0.57 & 0.075\\
~    & {\bf 2}& 0.0049 & 1.227 & 4.33 &  26.09  & 0.51 & 0.166\\
\hline
~    & 1      & 0      & 1.252 & 2.10 &  40.54  & 0.59 & 0.052 \\
1000 &{\bf 2} & 0.0039 & 1.051 & 4.54 &  36.42  & 0.51 & 0.125 \\
~    & 3      & 0.0038 & 0.979 & 6.24 &  33.50  & 0.51 & 0.186 \\
\hline
~    & 1      & 0      & 1.152 & 2.16 &  57.55  & 0.60 & 0.038 \\
~    & 2      & 0.0026 & 0.930 & 4.45 &  50.40  & 0.52 & 0.088 \\
2000 &{\bf 3} & 0.0008 & 0.818 & 6.80 &  44.98  & 0.50 & 0.151 \\
\hline
~    & 1      & 0      & 1.070 & 2.17 &  92.48  & 0.58 & 0.023\\
~    & 2      & 0.0005 & 0.897 & 4.49 &  72.01  & 0.55 & 0.062\\
4000 & 3      &-0.0014 & 0.758 & 6.80 &  65.84  & 0.51 & 0.103\\
~    &{\bf 4} &-0.0025 & 0.694 & 9.01 &  62.92  & 0.49 & 0.143\\
~    & 5      &-0.0031 & 0.652 &10.44 &  60.11  & 0.49 & 0.174\\
\hline\hline
\end{tabular}
\end{center}
\end{table}
\begin{table}[tb]
\caption{
Dependence on the number of encapsulated neon atoms $N_g$ and the number of layers $l$ of the relative specific energy $\Delta E$, pressure $P$, height $H$, radius $R$, and density $d$ of the stationary state of the nanobubble.
\label{tab4}
}
\begin{center}
\begin{tabular}{c|ccccccc}
\hline\hline
$N_g$& ~~$l$~&$\Delta E$~(eV)& $P$~(GPa) & $H$~(\AA) &  $R$~(\AA) &  $d$~(g/cm$^3$) & $H/R$\\
\hline
125  &{\bf 1} & 0      & 1.736 & 2.52 &  15.81  & 1.69 & 0.159\\
\hline
250  &{\bf 1} &   0    & 1.450 & 2.51 &  22.89  & 1.80 & 0.110\\
~    & 2      & 0.0101 & 1.476 & 4.64 &  21.03  & 1.71 & 0.221\\
\hline
500  & 1      & 0      & 1.224 & 2.50 &  32.71  & 1.86 & 0.076\\
~    &{\bf 2} & 0.0077 & 1.237 & 4.93 &  28.55  & 1.77 & 0.173\\
\hline
~    & 1      & 0      & 1.098 & 2.51 &  46.06  & 1.94 & 0.054 \\
1000 &{\bf 2} & 0.0034 & 1.007 & 4.92 &  38.17  & 1.83 & 0.129 \\
~    & 3      & 0.0068 & 1.028 & 7.16 &  36.39  & 1.80 & 0.197 \\
\hline
~    & 1      & 0      & 1.001 & 2.50 &  65.37  & 1.97 & 0.038 \\
~    & 2      & 0.0022 & 0.839 & 4.93 &  52.07  & 1.87 & 0.095 \\
2000 &{\bf 3} & 0.0034 & 0.872 & 7.40 &  48.46  & 1.83 & 0.153 \\
~    & 4      & 0.0050 & 0.876 & 9.25 &  46.98  & 1.82 & 0.197 \\
\hline
~    & 1      & 0      & 0.932 & 2.50 &  92.48  & 2.01 & 0.027\\
~    & 2      & 0.0011 & 0.727 & 4.93 &  72.01  & 1.90 & 0.068\\
4000 & 3      & 0.0015 & 0.760 & 7.38 &  65.84  & 1.85 & 0.112\\
~    &{\bf 4} & 0.0021 & 0.774 & 9.86 &  62.92  & 1.82 & 0.157\\
~    & 5      & 0.0030 & 0.752 &11.86 &  60.11  & 1.82 & 0.197\\
\hline\hline
\end{tabular}
\end{center}
\end{table}

It is convenient to define the base radius of the bubble as
$$
R = \frac{1}{N_r} \sum_{n=1\atop \Delta z_n \in [h_1, h_2]}^{N_c} r_n, \quad
N_r = \sum_{n=1 \atop \Delta z_n \in [h_1, h_2]}^{N_c} 1.
$$
Here, the distance to the center is 
$$
r_n = [(x_n^0 - x_c)^2 + (y_n^0 - y_c)^2]^{1/2},
$$
and the second vertical displacement threshold $h_2$ is 1.62, 1.95, 2.5, 2.7, and 3.1~\AA~
for He, Ne, Ar, Kr, and Xe atoms, respectively.

The volume of the encapsulation cavity beneath the graphene sheet (the nanobubble volume) is
$$
V = \sum_{n=1}^{N_c} \Delta z_n S_n^0,
$$
where $S_n^0$ is the area of the projection onto the plane $z = 0$ of the triangle corresponding to the $n$-th sheet atom (see the graphene sheet triangulation scheme in Fig.~\ref{fg01}).
The density of gas atoms inside the nanobubble is $d = M_g N_g / V$.

In the stationary state, a force acts on the $n$-th atom of the graphene sheet from the side of the encapsulated atoms:
\[
{\bf F}_n=\frac{24\epsilon_0}{\sigma^2}\sum_{k=N_c+1}^N\left[2(\sigma/r_{nk})^{14}-(\sigma/r_{nk})^8\right]
({\bf u}_n^0-{\bf u}_k^0),
\]
where the distance between atoms is $r_{nk}=|{\bf u}_n^0-{\bf u}_k^0|$.
Consequently, the local pressure on this atom is $p_n=({\bf F}_n,{\bf e}_n)/S_n$, where ${\bf e}_n$ is the unit vector orthogonal to the plane of the triangle containing the sheet atom, and $S_n$ is the area of this triangle (see Fig. \ref{fg01}).
The value of the average pressure on the external surface of the nanobubble is
\[
P=\frac{1}{N_i}\sum_{n=1\atop \Delta z_n > h_1}^{N_c} p_n.
\]
This value also determines the average value of the internal pressure.

\begin{table}[tb]
\caption{
Dependence on the number of encapsulated argon atoms $N_g$ and the number of layers $l$ of the relative specific energy $\Delta E$, pressure $P$, height $H$, radius $R$, and density $d$ of the stationary state of the nanobubble.
\label{tab5}
}
\begin{center}
\begin{tabular}{c|ccccccc}
\hline\hline
$N_g$& ~~$l$~&$\Delta E$~(eV)& $P$~(GPa) & $H$~(\AA) &  $R$~(\AA) &  $d$~(g/cm$^3$) & $H/R$\\
\hline
125  &{\bf 1}     & 0      & 1.601 & 3.24 &  20.65  & 1.64 & 0.157\\
\hline
250  & 1      &   0    & 1.329 & 3.23 &  29.33  & 1.75 & 0.110\\
~    &{\bf 2} & 0.0156 & 1.446 & 6.09 &  26.01  & 1.68 & 0.234\\
\hline
500  & 1      & 0      & 1.159 & 3.23 &  41.22  & 1.82 & 0.078\\
~    &{\bf 2} & 0.0045 & 1.231 & 6.22 &  34.60  & 1.77 & 0.180\\
\hline
~    & 1      & 0      & 1.032 & 3.23 &  58.35  & 1.87 & 0.055 \\
1000 & 2      &-0.0028 & 1.067 & 6.21 &  46.94  & 1.82 & 0.132 \\
~    &{\bf 3} & 0.0036 & 1.098 & 9.16 &  44.54  & 1.79 & 0.206 \\
\hline
~    & 1      & 0      & 0.946 & 3.23 &  82.38  & 1.91 & 0.039 \\
~    & 2      &-0.0061 & 0.939 & 6.22 &  64.20  & 1.87 & 0.097 \\
2000 &{\bf 3} &-0.0045 & 0.974 & 9.22 &  58.45  & 1.83 & 0.158 \\
~    & 4      &-0.0006 & 1.007 &12.04 &  56.71  & 1.81 & 0.212 \\
\hline
~    & 1      & 0      & 0.886 & 3.23 & 116.39  & 1.94 & 0.028\\
~    & 2      &-0.0089 & 0.845 & 6.22 &  88.35  & 1.90 & 0.070\\
4000 & 3      &-0.0098 & 0.899 & 9.22 &  80.29  & 1.86 & 0.115\\
~    &{\bf 4}  &-0.0085 & 0.919 &12.24 &  76.26  & 1.84 & 0.161\\
~    & 5      &-0.0060 & 0.920 &15.09 &  72.65  & 1.83 & 0.208\\
\hline\hline
\end{tabular}
\end{center}
\end{table}
\begin{table}[tb]
\caption{
Dependence on the number of encapsulated krypton atoms $N_g$ and the number of layers $l$ of the relative specific energy $\Delta E$, pressure $P$, height $H$, radius $R$, and density $d$ of the stationary state of the nanobubble.
\label{tab6}
}
\begin{center}
\begin{tabular}{c|ccccccc}
\hline\hline
$N_g$& ~~$l$~&$\Delta E$~(eV)& $P$~(GPa) & $H$~(\AA) &  $R$~(\AA) &  $d$~(g/cm$^3$) & $H/R$\\
\hline
125  &{\bf 1} &  0     & 1.599   & 3.48  &  22.31  &  2.80 & 0.156\\
~    & 2 & 0.0296 & 1.714   & 5.72  &  21.28  &  2.67 & 0.269\\
\hline
250  & 1 & 0      & 1.368   & 3.48  &  31.44  &  2.97 & 0.182\\
~    &{\bf 2} & 0.0179 & 1.480   & 6.57  &  27.61  &  2.87 & 0.238\\
\hline
500  & 1 & 0      & 1.200   & 3.48  &  44.27  &  3.10 & 0.079\\
~    &{\bf 2}& 0.0043 & 1.293   & 6.64  &  37.07  &  3.01 & 0.179\\
\hline
~    & 1 & 0      & 1.099   & 3.48  &  62.51  &  3.18 & 0.056\\
1000 & 2 &-0.0052 & 1.148   & 6.64  &  50.02  &  3.12 & 0.133\\
~    &{\bf 3}& 0.0032 & 1.162   & 9.79  &  46.38  &  3.08 & 0.211\\
\hline
~    & 1 & 0      & 1.021   & 3.48  &  88.31  &  3.25 & 0.039\\
2000 & 2 &-0.0095 & 1.035   & 6.64  &  68.59  &  3.20 & 0.097\\
~    &{\bf 3}&-0.0073 & 1.068   & 9.82  &  62.33  &  3.15 & 0.158\\
~    & 4 & 0.0024 & 1.085   & 12.88 &  60.19  &  3.12 & 0.214\\
\hline
~    & 1 & 0      & 0.968   & 3.48  &  124.55 &  3.30 & 0.028\\
~    & 2 & -0.0128& 0.951   & 6.64  &  94.53  &  3.26 & 0.070\\
4000 & 3 & -0.0134& 0.997   & 9.82  &  85.73  &  3.20 & 0.115\\
~    &{\bf 4}& -0.0116& 1.019   & 13.02 &  81.56  &  3.17 & 0.160\\
~    & 5 & -0.0080& 1.025   & 16.09 &  77.39  &  3.16 & 0.208\\
\hline\hline
\end{tabular}
\end{center}
\end{table}

Solving problem (\ref{f12}) reveals that each stationary state of a nanobubble possesses a strictly defined number of atomic layers, $l$, within the internal cluster---see Figs. \ref{fg02} and \ref{fg03}.
For systems with up to $N_g\le 4000$ encapsulated atoms, stationary states with $l=1,...,6$ layers are possible.
Each internal atom resides within one of these layers (Fig. \ref{fg03}). 
The layers are flat and circular in shape. 
They are concentrically stacked, forming a symmetric $l$-stepped pyramid with a flat top, and remain parallel to the substrate plane. 
While atoms within each layer lie in the same plane, the edge atoms are displaced downward, effectively smoothing the stepped profile of the pyramid.
The size of the inert gas atoms varies across the series, resulting in distinct interlayer distances: $\Delta h=2.2$, 2.5, 3.0, 3.2, and 3.6~\AA{} for He, Ne, Ar, Kr, and Xe, respectively. 
For argon atoms specifically, two- and three-layer structures at $N_g=2463$ were first reported in \cite{Iakovlev2017}.

Covering the $l$-stepped pyramid from above with a graphene sheet is achieved through its local stretching.
The valence bonds of the sheet stretch over the group of encapsulated atoms, while outside this zone the sheet remains in an undeformed state, lying flush against the substrate---see Fig. \ref{fg04}.
The most significant stretching of C--C bonds occurs over the layer boundaries, where the deformation energy can reach values of 0.01 eV per carbon atom.
Here, valence bonds can elongate by 1\%. 
This is in agreement with experimental measurements on graphene bubbles \cite{Georgiou2011, Zabel2011}.
As seen in Fig. \ref{fg04}, the zones of sheet stretching strictly correspond to the stepped structure of the pyramid of encapsulated atoms.
For example, for $N_g=4000$ and $l$-layer states with $l = 1, 2, 3$, stretching of the sheet mainly occurs above the edges of the layers.
For states with a larger number of layers ($l = 4, 5$), stretching occurs more uniformly across the entire surface of the nanobubble, but the maximum values are always reached at its base (over the edge of the first layer).

Local stretching of the valence bonds of the graphene sheet induces pressure $p_n$ on the encapsulated atoms.
As shown in Fig. \ref{fg05}, the pressure arising from encapsulation is localized only on the surface of the nanobubble.
Over the layer boundaries, the pressure can reach maximum values of up to 10 GPa, while the average pressure value is \( P \sim 1 \) GPa.
The highest pressure is experienced by single-layer packings of internal atoms.
The average pressure value decreases with an increasing number of layers.
\begin{table}[tb]
\caption{
Dependence on the number of encapsulated xenon atoms $N_g$ and the number of layers $l$ of the relative specific energy $\Delta E$, pressure $P$, height $H$, radius $R$, and density $d$ of the stationary state of the nanobubble.
\label{tab7}
}
\begin{center}
\begin{tabular}{c|ccccccc}
\hline\hline
$N_g$& ~~$l$~&$\Delta E$~(eV)& $P$~(GPa) & $H$~(\AA) &  $R$~(\AA) &  $d$~(g/cm$^3$) & $H/R$\\
\hline
125  &{\bf 1}&  0      & 1.545   & 4.01  &  25.39  &  2.99 &  0.158\\
~    & 2 &  0.0347 & 1.646   & 6.69  &  24.09  &  2.85 &  0.278\\
\hline
250  & 1 &  0      & 1.324   & 4.01  &  35.82  &  3.14 &  0.112\\
~    &{\bf 2}&  0.0134 & 1.411   & 7.54  &  31.48  &  3.04 &  0.240\\
\hline
500  & 1 &  0      & 1.182   & 4.02  &  50.36  &  3.27 & 0.080\\
~    &{\bf 2}& -0.0081 & 1.238   & 7.64  &  41.87  &  3.19 & 0.182\\
~    & 3 &  0.0087 & 1.265   &10.51  &  40.62  &  3.13 & 0.259\\
\hline
~    & 1 & 0      & 1.087   & 4.02  &  71.04  &  3.36 & 0.057\\
1000 & 2 & -0.0226& 1.113   & 7.64  &  56.73  &  3.30 & 0.135\\
~    &{\bf 3}& -0.0154& 1.141   & 11.23 &  52.79  &  3.25 & 0.213\\
\hline
~    & 1 & 0      & 1.021   & 4.02  & 100.34  &  3.42 & 0.040\\
2000 & 2 &-0.0298 & 1.012   & 7.64  &  77.41  &  3.38 & 0.099\\
~    &{\bf 3}&-0.0322 & 1.047   & 11.27 &  70.32  &  3.34 & 0.160\\
~    & 4 &-0.0269 & 1.060   & 14.74 &  68.33  &  3.30 & 0.216\\
\hline
~    & 1 & 0      & 0.976   & 4.02  &  141.59 &  3.48 & 0.028\\
~    & 2 & -0.0350& 0.948   & 7.65  &  107.08 &  3.44 & 0.071\\
4000 & 3 & -0.0411& 0.977   & 11.28 &  96.85  &  3.38 & 0.116\\
~    &{\bf 4}& -0.0412& 0.989   & 14.94 &  91.93  &  3.35 & 0.163\\
~    & 5 & -0.0380& 0.995   & 18.43 &  87.79  &  3.34 & 0.210\\
~    & 6 & -0.0345& 0.996   & 20.81 &  86.63  &  3.33 & 0.240\\
\hline\hline
\end{tabular}
\end{center}
\end{table}
\begin{table}[tb]
\caption{
Dependence on the adhesion energy $\varepsilon_0$ and the number of layers $l$ of the relative specific energy $\Delta E$, pressure $P$, height $H$, radius $R$, and density $d$ of the stationary state of the nanobubble with $N_g=4000$ argon atoms.
\label{tab8}
}
\begin{center}
\begin{tabular}{c|ccccccc}
\hline\hline
$\varepsilon_0$~(eV)& ~~$l$~&$\Delta E$~(eV)& $P$~(GPa) & $H$~(\AA) &  $R$~(\AA) &  $d$~(g/cm$^3$) & $H/R$\\
\hline
~    & 1 & 0      & 0.684   & 3.27  &  119.76 &  1.81 & 0.027\\
~    & 2 & 0.0165 & 0.709   & 6.27  &   92.22 &  1.78 & 0.068\\
0.037&{\bf 3}& 0.0035 & 0.729   & 9.27  &   83.28 &  1.76 & 0.111\\
~    & 4 & 0.0061 & 0.728   & 12.30 &  79.03  &  1.75 & 0.156\\
~    & 5 & 0.0095 & 0.727   & 15.00 &  75.83  &  1.74 & 0.198\\
\hline
~    & 1 & 0      & 1.407   & 3.17  &  112.57 &  2.12 & 0.028\\
~    & 2 &-0.0247 & 1.118   & 6.14  &   85.36 &  2.05 & 0.072\\
0.148& 3 &-0.0302 & 1.209   & 9.14  &   77.63 &  2.00 & 0.118\\
~    & 4 &-0.0317 & 1.241   & 12.18 &  73.43  &  1.96 & 0.166\\
~    &{\bf 5}&-0.0310 & 1.249   & 15.08 &  69.62  &  1.95 & 0.217\\
\hline\hline
\end{tabular}
\end{center}
\end{table}

The maximum possible number of layers, $l_m$, depends on the number of encapsulated atoms, $N_g$, increasing monotonically with $N_g$.

For helium atoms, the following behavior is observed:
for $N_g<N_1=188$, only single-layer packings are possible ($l_m = 1$);
for $N_1<N_g<N_2=850$, both single- and double-layer packings are possible ($l_m = 2$);
for $N_2< N_g< N_3=2500$, the maximum number of layers is $l_m = 3$;
for $N_3<N_g<N_4 = 3500$, it is $l_m = 4$;
and for $N_4<N_g<4000$, it is $l_m = 5$.
The boundary values of the number of atoms, $N_k$ for the other noble gas are as follows:
$N_k=182$, 650, 1500, 3500 for Ne; $N_k=145$, 550, 1600, 2500 for Ar;
$N_k=123$, 450, 1300, 2400 for Kr; and $N_k=113$, 430, 1200, 2200 for Xe.

A single-layer packing exists for any number of atoms.
Therefore, the relative energy of a stationary state can be measured from the energy of the single-layer packing: \( \Delta E = (E_l - E_1) / N_g \), where \( E_l \) is the energy of the \( l \)-layer stationary state.

The dependence of the relative specific energy $\Delta E$, internal pressure $P$, height $H$, radius $R$, and density $d$ of the nanobubble's stationary state on the number of helium atoms $N_g$ is given in Table~\ref{tab3}.
As can be seen from the table, single-layer stationary states are the most energetically favorable for $N_g \le 2000$.
The pressure inside the encapsulation monotonically decreases with an increasing number of atoms.
With an increasing number of layers, a slight decrease in pressure and density is observed.
The height of the nanobubble, $H$, always takes a discrete set of values proportional to the number of layers $l$: $H \approx l\Delta h$.

For neon, argon, krypton and xenon atoms, the dependencies of $\Delta E$, $P$, $H$, $R$, and $d$ on the number of encapsulated atoms $N_g$ are presented in Tables~\ref{tab4}, \ref{tab5}, \ref{tab6} and \ref{tab7}, respectively.
As can be seen from the tables, for all encapsulated atoms, the pressure inside the nanobubble decreases  monotonically with an increasing number of atoms.
Pressure and density reach their maximum values for single-layer states.

For neon atoms, single-layer stationary states are the most energetically favorable across the entire range investigated ($N_g \le 4000$).
For argon atoms, single-layer stationary states are the most energetically favorable only for $N_g < 800$.
For $800 < N_g \le 3000$, two-layer states become more favorable, and for $N_g > 3000$, three-layer states are preferred.
For krypton atoms, single-layer stationary states are the most energetically favorable for $N_g < 700$. Two-layer states are favorable for $700 < N_g < 3000$, and three-layer states are favored for $N_g > 3000$.
For xenon atoms, single-layer stationary states are the most energetically favorable for $N_g < 500$, while two-layer states are favorable for $500 < N_g < 2000$.
For $N_g > 2000$, three- and four-layer states become increasingly favorable.
\begin{figure}[tb]
\begin{center}
\includegraphics[angle=0, width=1.0\linewidth]{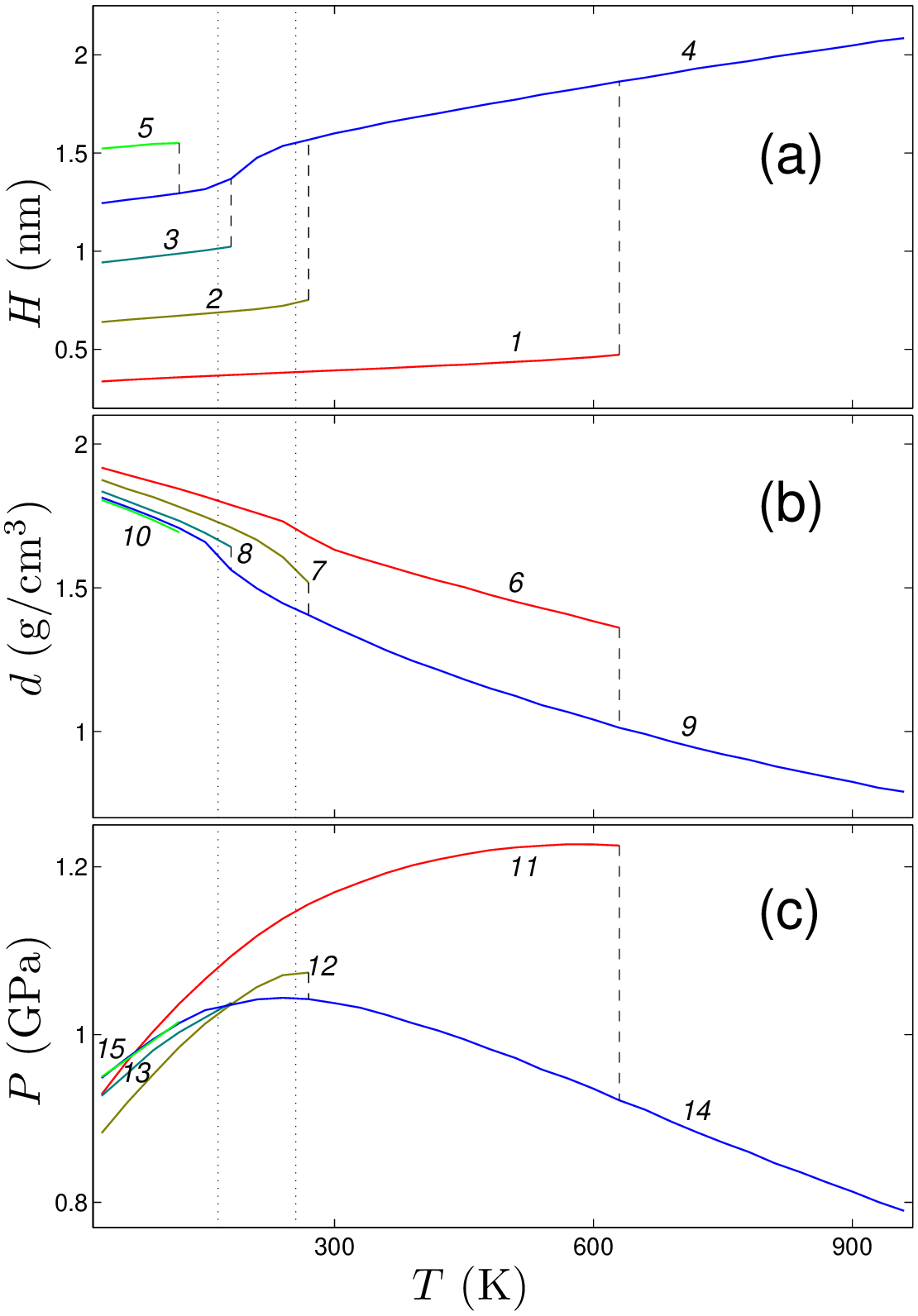}
\end{center}
\caption{\label{fg06}\protect
Dependence on temperature $T$ of (a) height $H$, (b) density $d$, and (c) internal pressure $P$ for a graphene nanobubble with $N_g = 4000$ argon atoms.
Curves 1, 6, 11 correspond to the state with $l = 1$; curves 2, 7, 12 -- to $l = 2$, ..., and curves 5, 10, 15 -- to $l = 5$ layers.
The dashed vertical lines correspond to the transition temperatures between states $l \rightarrow 4$: $T_1 = 630$, $T_2 = 270$, $T_3 = 180$, and $T_5 = 120$ K.
The dotted vertical lines correspond to the melting temperature of the single-layer ($T_{1,m}=255$~K) and four-layer state ($T_{4,m}=160$~K).
}
\end{figure}
\begin{figure}[tb]
\begin{center}
\includegraphics[angle=0, width=1.0\linewidth]{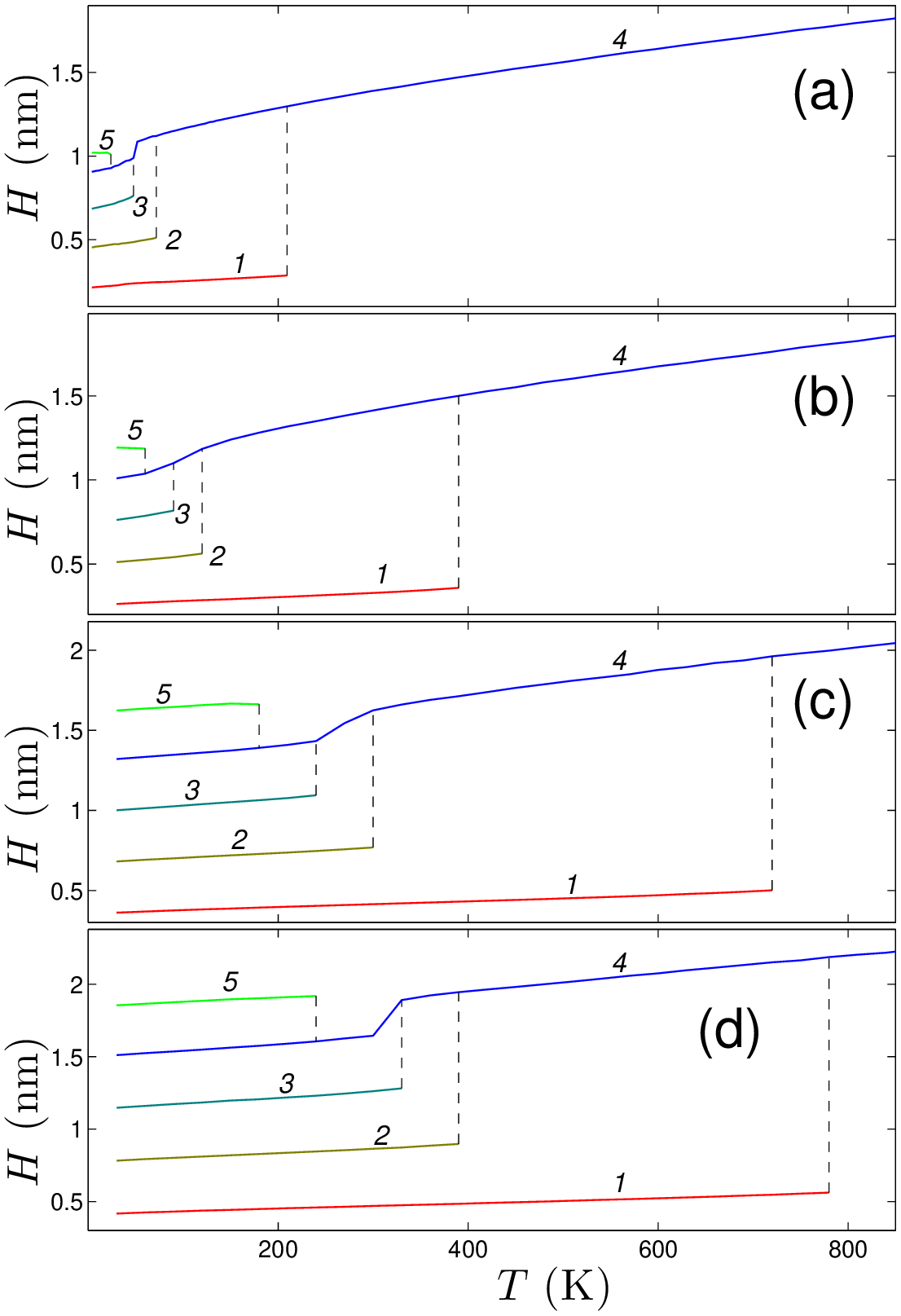}
\end{center}
\caption{\label{fg07}\protect
Dependence on temperature $T$ of height of graphene nanobubble $H$ with $N_g=4000$ internal
(a) helium, (b) neon, (c) krypton and (d) xenon atoms (curves 1,..., 5 correspond to the state
with $l=1$,..., 5 layers.
}
\end{figure}
\begin{figure}[tb]
\begin{center}
\includegraphics[angle=0, width=1.0\linewidth]{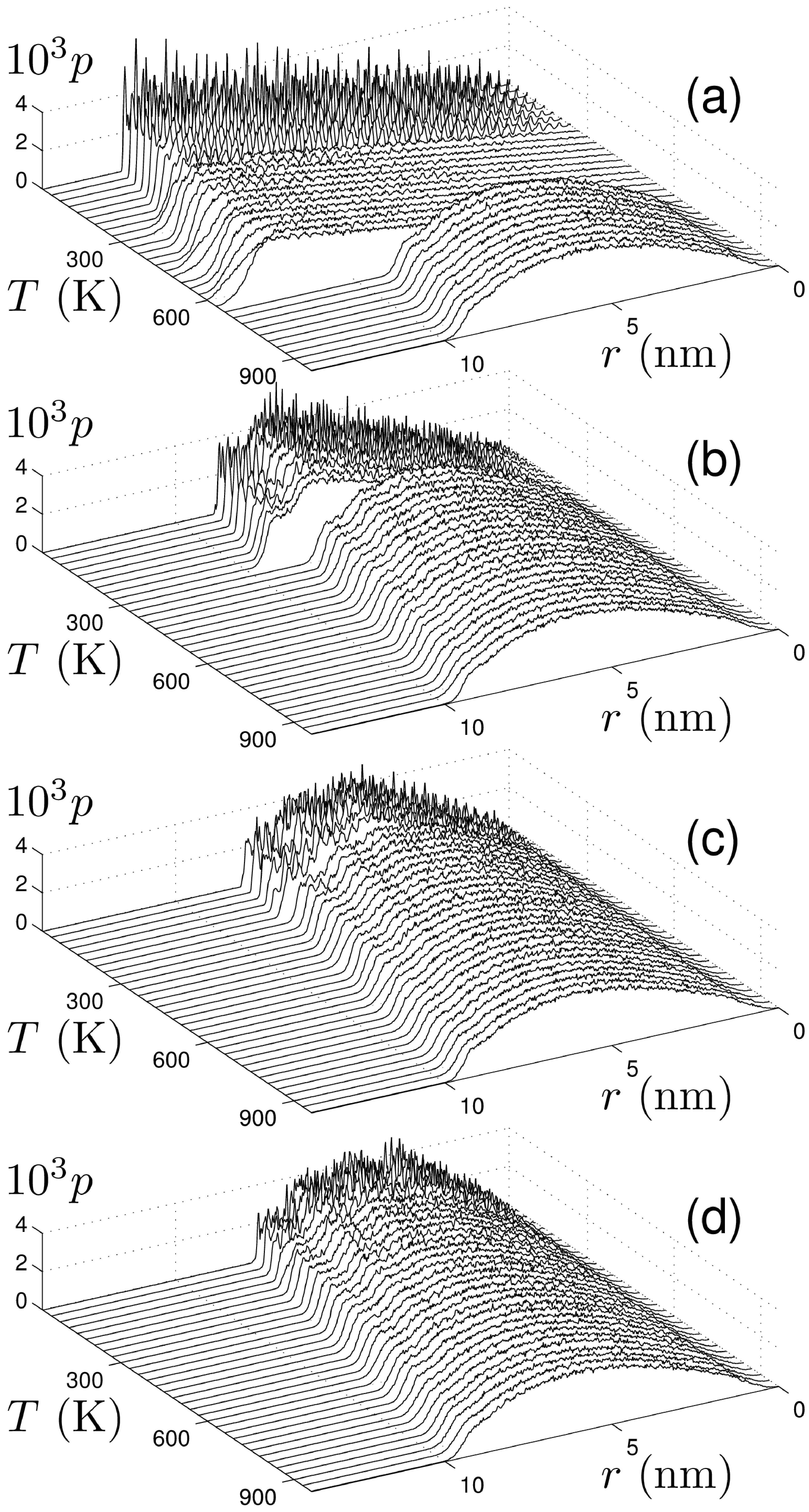}
\end{center}
\caption{\label{fg08}\protect
Dependence on temperature $T$ of the vertical displacement distribution $p(z)$ for $l$-layer states of $N_g = 4000$ internal argon atoms: (a) $l = 1$; (b) $l = 2$; (c) $l = 3$; and (d) $l = 4$.
}
\end{figure}

To understand how the shape of a nanobubble's stationary state depends on the value of the interaction energy between the graphene sheet and the substrate, we also consider states with twice weaker and twice stronger interactions: at $\epsilon_0=0.037$ and $0.148$~eV.
The weak interaction is characteristic of substrates with low density. 
For example, for an I$_h$ ice substrate, the energy is $\epsilon_0=0.029$~eV \cite{Savin2019}. 
The strongest interaction occurs with the surface of a nickel crystal [with Ni(111) surface], where $\varepsilon_0=0.133$~eV \cite{Lahiri2011}. 
Thus, the three values of adhesion energy $\varepsilon=0.037$, 0.074, and 0.148~eV correspond to weak, medium (the most typical), and strong interaction of the graphene sheet with the substrate.

The numerical solution of the problem (\ref{f12}) showed that for all three characteristic values of the adhesion energy, a nanobubble containing $N_g=4000$ argon atoms can only exist in stationary states with $l=1,...,5$ layers; i.e., the maximum possible number of layers is $l_m=5$.
The dependence of $\Delta E$, $P$, $H$, $R$, and $d$ on the number of layers is presented in Tables \ref{tab5} and \ref{tab8}. 
As can be seen from the tables, the shape of the nanobubble's stationary state depends only weakly on the value of $\epsilon_0$. 
A twofold increase in the adhesion energy leads to an increase in the internal pressure $P$, accompanied by a slight decrease in the nanobubble radius $R$.
With weak adhesion, the most energetically favorable state is a single-layer nanobubble; with medium adhesion, a three-layer state; and with strong adhesion, a five-layer stationary state is the most favorable. Thus, an increase in the interaction energy between the graphene sheet and the substrate does not significantly change the step-like shape of the stationary states, but it noticeably increases the internal pressure and makes multilayer states energetically more favorable.

\section{Stability of Stationary States to Thermal Fluctuations  \label{sec4}}
Note that all stationary states of nanobubbles obtained by solving the energy minimization problem are always stable.
To verify their stability against thermal fluctuations, we performed simulations of nanobubble dynamics at temperatures $T\le 1000$~K.
\begin{figure}[tb]
\begin{center}
\includegraphics[angle=0, width=1.0\linewidth]{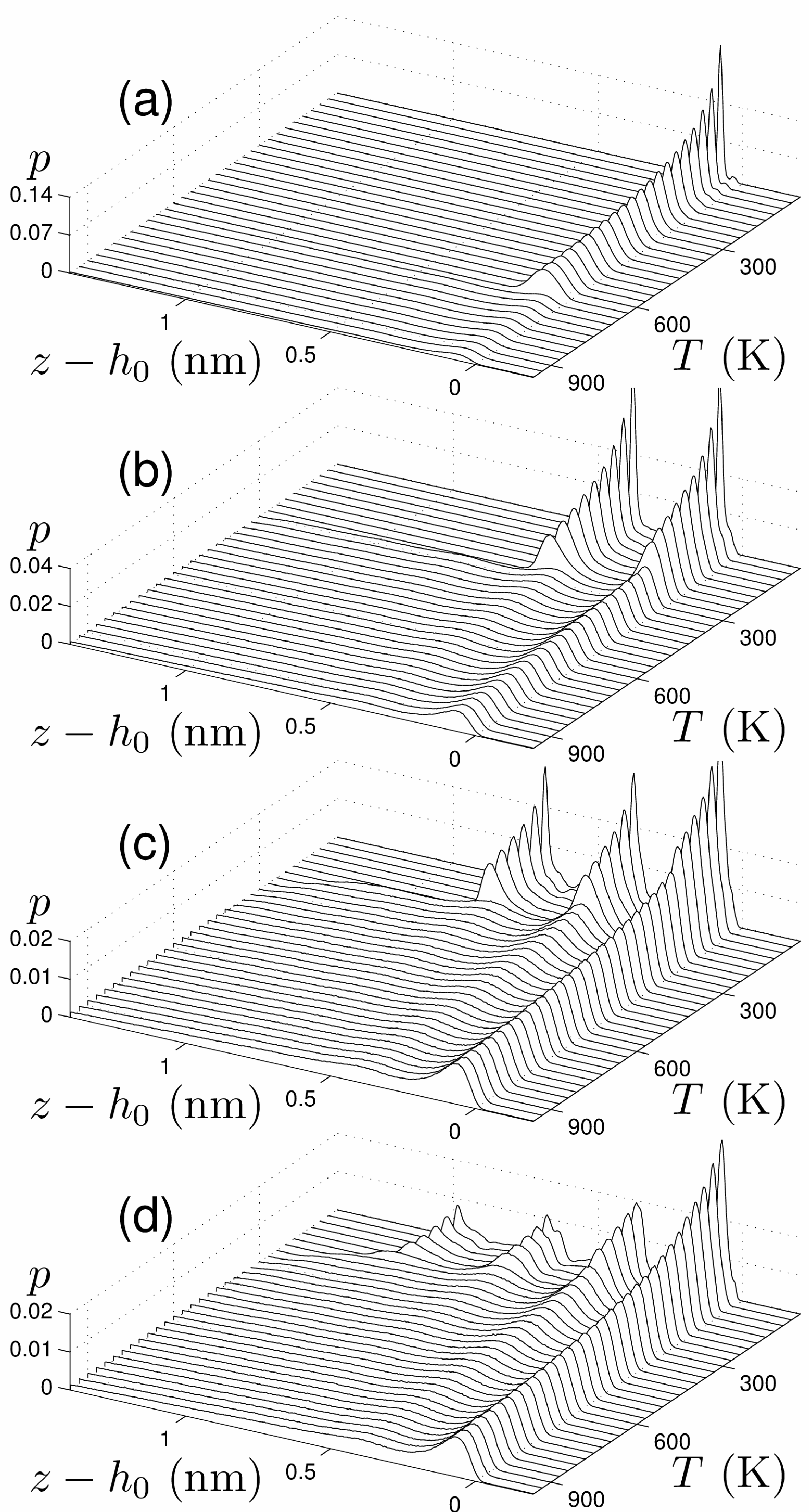}
\end{center}
\caption{\label{fg09}\protect
Temperature dependence $T$ of the transverse displacement distribution $p(r)$ from the center of mass for $l$-layer states of $N_g=4000$ inner argon atoms: (a) $l=1$; (b) $l=2$; (c) $l=3$; and (d) $l=4$.
}
\end{figure}

Let us consider the dynamics of a molecular system in which a graphene sheet interacts with a Langevin thermostat.
Dynamics of a thermalized graphene sheet with encapsulated atoms is described by the system of Langevin equations:
\begin{eqnarray}
M_n \ddot{\bf u}_n &=& -\frac{\partial H}{\partial {\bf u}_n} - \Gamma M_n \dot{\bf u}_n + \Xi_n,~~n=1,...,N_c,~~~
\label{f13}\\
M_n \ddot{\bf u}_n &=& -\frac{\partial H}{\partial {\bf u}_n},~~n=N_c+1,...,N,
\label{f14}
\end{eqnarray}
where the mass $M_n = M_c$ for $n=1,...,N_c$ and $M_n = M_g$ for $n=N_c+1,..., N$ (the total number of atoms is $N=N_c+N_g$).
Here, $\Gamma = 1/t_r$ is the friction coefficient characterizing the intensity of energy exchange with the thermostat (relaxation time $t_r = 10$~ps), and $\Xi_n=\{\xi_{n,i}\}_{i=1}^3$ is a three-dimensional vector of normally distributed random forces normalized by the following conditions:
$$
\langle\xi_{n,i}(t)\xi_{k,j}(s)\rangle = 2\Gamma M_n k_B T \delta_{nk}\delta_{ij}\delta(t-s).
$$
Here, $T$ is the thermostat temperature and $k_B$ is the Boltzmann constant.

The relaxation time $t_r$ determines the intensity of energy exchange with the thermostat.
In the simulated molecular system, the role of the thermostat is played by a flat substrate, to which the atoms are weakly coupled via non-valence interactions.
For such interactions, the relaxation time is typically $t_r \sim 100$~ps.
For the convenience of numerical simulation, we use a smaller value, $t_r = 10$~ps.
This allows us to significantly reduce the numerical integration time required to reach an equilibrium state and obtain reliable averages.

We numerically integrate the system of equations of motion (\ref{f13}), (\ref{f14}) with initial conditions corresponding to a stationary nanobubble state:
\begin{equation}
\{ {\bf u}_n(0)={\bf u}_n^0,~\dot{\bf u}_n(0)=0\}_{n=1}^N.
\label{f15}
\end{equation}
The equations of motion Eq. (\ref{f13}), (\ref{f14}) with initial conditions (\ref{f15}) are solved numerically using the velocity Verlet method \cite{Verlet1967}.
A time step of 1 fs is used in the simulations, as further reduction of the time step has no appreciable effect on the results.

Numerical simulations of the dynamics showed that among the $l$-layer states of the nanobubble, there is always one "ground"\ state, which smoothly transitions into a layerless liquid configuration as the temperature increases.
All other states transition into this ground state at a characteristic temperature $T_l$.
The dependence of the nanobubble's height $H$, density $d$, and internal pressure $P$ on temperature is shown in Fig. \ref{fg06}.

As can be seen from the figure, for a bubble with $N_g=4000$ argon atoms, a single-layer packing ($l=1$) persists up to a very high temperature $T_1 = 630$~K.
Further temperature increase leads to its transition into a layerless molten configuration, accompanied by a sharp increase in height $H$, a decrease in density $d$, and internal pressure $P$.
Two-layer atom packings ($l=2$) persist up to a temperature $T_2 =270$~K, at which they transition into a four-layer packing.
This transition is also accompanied by a sharp increase in height and a decrease in density.
Three-layer packings ($l=3$) persist up to a temperature of $T_3 = 180$~K, and five-layer packings ($l=5$) persist up to $T_5 = 120$~K (see Fig. \ref{fg06} (a)).
Here, the ground state is the four-layer $(l=4)$ packing.
Note that with weak adhesion $\varepsilon_0=0.037$~eV, the ground state is the three-layer ($l=3$) state, while with strong adhesion $\varepsilon_0=0.148$~eV, it is the five-layer ($l=5$) state.

A similar pattern is observed for other noble gases.
For He, the transition temperatures are $T_l=209$, 72, 47, and 23~K;
for Ne---$T_l =390$, 120, 90, and 60~K;
for Kr---$T_l=720$, 300, 240, and 180~K;
for Xe---$T_l=780$, 390, 330, and 240~K (where $l = 1$, 2, 3, and 5, respectively), as shown in Fig.~\ref{fg07}.

In Tables \ref{tab3} to \ref{tab8}, the number of layers $l$ corresponding to the ground states is highlighted in bold.
Note that the ground states are not always the most energetically favorable.
Their key feature is the smooth (non-abrupt) change in their shape with increasing temperature.
The ground state can also be obtained from the layerless configuration of the nanobubble at $T = 900$~K by gradually reducing the thermostat temperature to zero.

Conformational changes of the encapsulated atom cluster are reflected in the change in the distribution of atoms by their vertical $\{z_n\}_{n=N_c}^N$ (Fig. \ref{fg08}) and radial (in-plane) displacements from the center of mass:
$$
\{r_n = [(x_n - x_c)^2 + (y_n - y_c)^2]^{1/2}\}_{n=N_c+1}^N
$$
(Figs. \ref{fg09}).

As can be seen from Fig.~\ref{fg08} (a,b,c), the shape of the vertical displacement distribution changes sharply at the temperature of $l$-layer state stability loss.
For $l \neq 4$, the clearly pronounced layered structure of the cluster disappears here.
For the ground state ($l=4$), only a gradual, smooth broadening of the layered structure occurs with increasing temperature, with only the first layer of atoms adjacent to the substrate remaining clearly pronounced---see Fig.~\ref{fg08} (d).

Radial displacements also show changes in the distribution shape at the transition temperature $T_l$.
The transition to a state with a larger number of layers leads to a sharp decrease in the diameter of the inner atom cluster---see Fig.~\ref{fg09}~(a,b,c).
For the ground state with $l=4$ layers, only a continuous, gradual increase in the cluster diameter (its thermal expansion) occurs---see Fig.~\ref{fg09}~(d).

The shape of the radial distribution also allows one to judge the phase state of the cluster.
At low temperature, the radial distribution has a discrete structure characteristic of a crystalline (solid) state.
Upon reaching the melting temperature $T_{l,m}$, the discrete distribution shape is replaced by a continuous one, characteristic of a liquid state.
For $N_g = 4000$ argon atoms, the melting of the single-layer cluster occurs at $T_{1,m}\in$(240,270), and of the four-layer cluster at $T_{4,m}\in(150, 180)$~K---see Fig. \ref{fg09}.
For helium atoms, the melting temperatures are $T_{1,m}\approx 80$, $T_{4,m}\approx 48$,
for neon $T_{1,m}\approx 135$, $T_{4,m}\approx 75$,
for krypton $T_{1,m}\approx 315$, $T_{4,m}\approx 240$,
and for xenon $T_{1,m}\approx 435$, $T_{4,m}\approx 285$~K.
\begin{figure}[tb]
\begin{center}
\includegraphics[angle=0, width=1.0\linewidth]{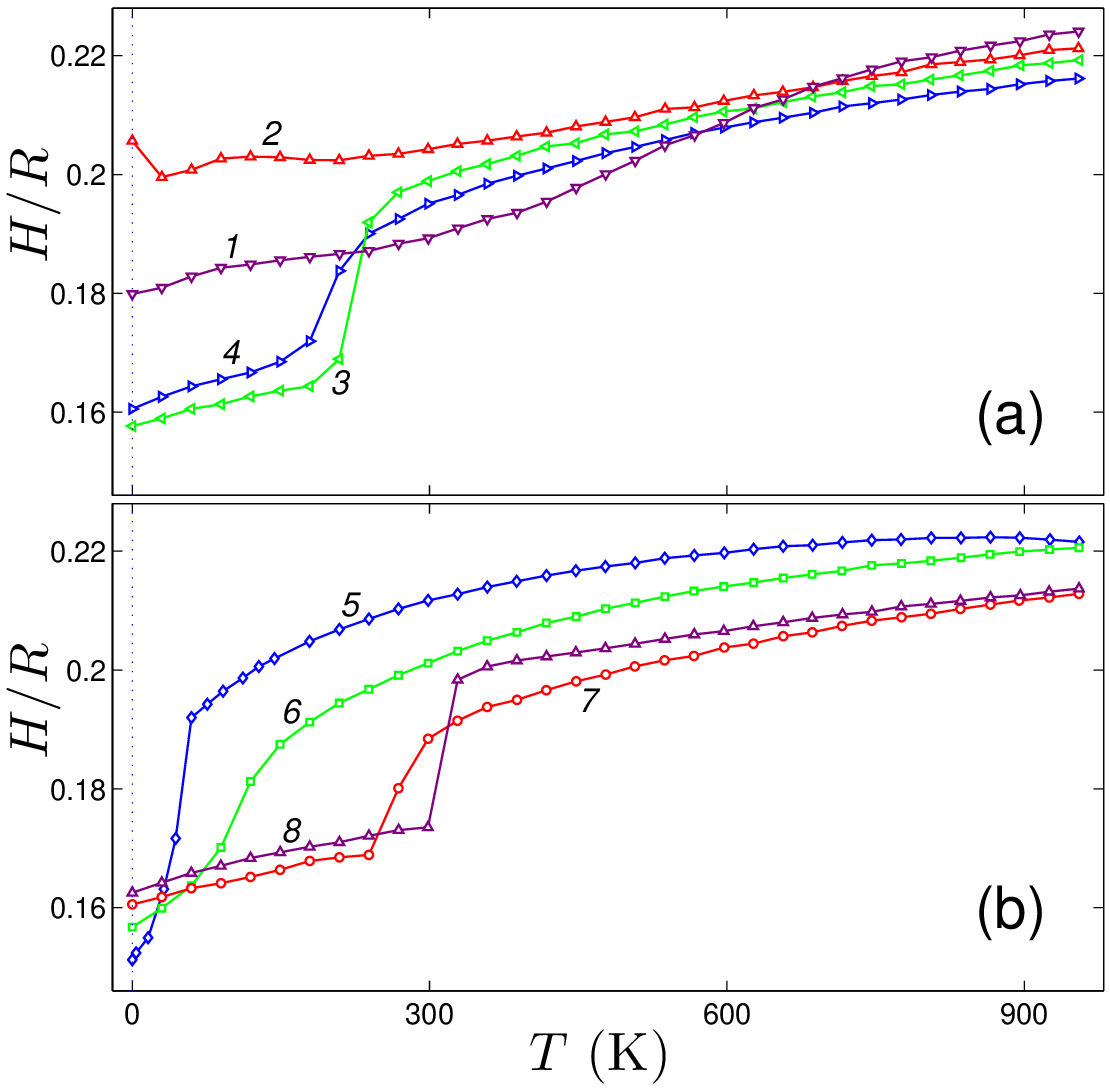}
\end{center}
\caption{\label{fg10}\protect
Temperature dependence $T$ of the ratio $H/R$ for the ground state of the nanobubble
with (a) $N_g=500$, 1000, 2000, 4000 argon atoms (curves 1, 2, 3, 4)
and (b) with $N_g=4000$ helium, neon, krypton, xenon atoms (curve 5, 6, 7, 8).
}
\end{figure}
\begin{figure}[tb]
\begin{center}
\includegraphics[angle=0, width=1.0\linewidth]{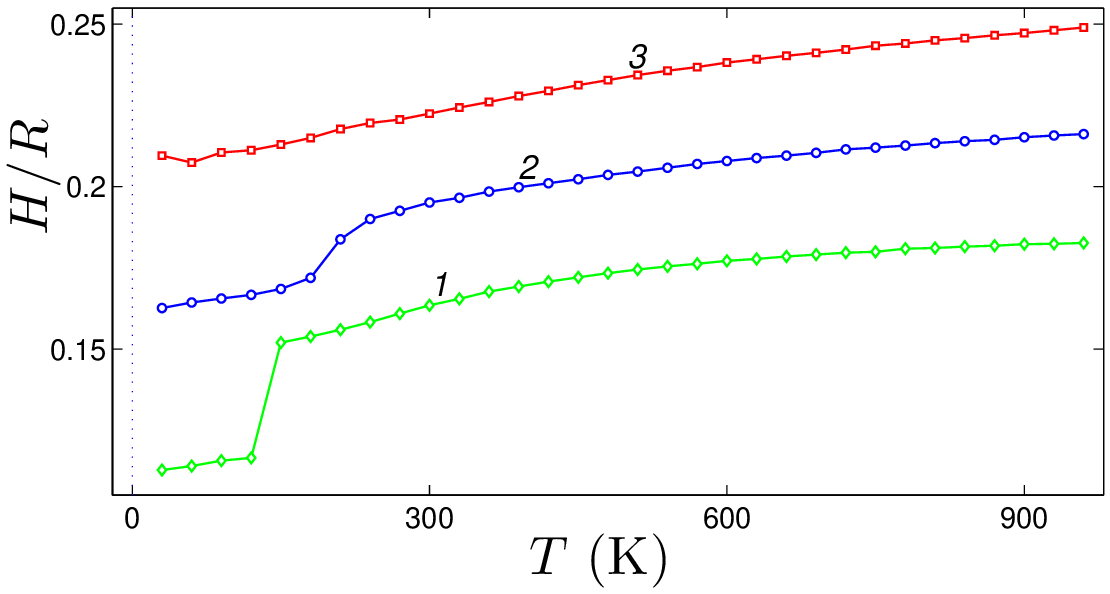}
\end{center}
\caption{\label{fg11}\protect
Temperature dependence $T$ of the ratio $H/R$ for the ground state of the nanobubble
with $N_g=4000$ argon atoms for adhesion $\epsilon_0=0.037$, 0.074, 0.148~eV (curves 1, 2, 3)
}
\end{figure}

The obtained values of the one-layer stability temperature and the ground state melting temperature do not depend on the number of atoms in the encapsulated cluster; the same values are obtained for $N_g=500$, 1000, 2000, and 3000.
Therefore, it can be expected that the obtained estimates will also hold for larger clusters (for $N_g> 4000$).
Noble gases transition from the solid to the liquid state at very low temperatures (24, 84, 114, and 161~K for Ne, Ar, Kr, and Xe, respectively).
The high melting temperatures of their encapsulated clusters are explained by the presence of high internal pressure and the effect of strong spatial confinement \cite{Faraji2022}.

The performed simulations show that the encapsulated noble gas cluster (graphene nanobubble) ceases to be a multistable system only at high temperatures $T>T_1$.
Thus, at room temperature ($T=300$~K), graphene nanobubbles of Ne, Ar, Kr, and Xe, in addition to the ground three-dimensional state, also have a stable quasi-two-dimensional one-layer state---see Figs.~\ref{fg06} (a) and \ref{fg07} (b, c, d).

The existence of several stable states with different numbers of atomic layers $l$ at low temperatures indicates that nanobubbles in this regime cannot possess a universal (scalable) shape; consequently, the height-to-radius ratio is not constant.
Indeed, for a single-layer state ($l=1$), increasing the number of encapsulated atoms $N_g$ leaves the nanobubble's height unchanged ($H = 2.2$, 2.5, 3.2, 3.5, and 4.0~\AA~for He, Ne, Ar, Kr, and Xe, respectively), while its radius $R$ grows as $\sqrt{N_g}$.
The ratio $H/R$ therefore approaches zero as $N_g \rightarrow \infty$.

Based on the data in Tables~\ref{tab3} to~\ref{tab8}, the ratio $H/R$ can vary from 0 to 0.28 here. The violation of the scaling rule for graphene bubbles with a radius of $\sim 1$ nm was noted in the experimental study \cite{Villarreal2021}, where they were created via the implantation of noble gas ions (He, Ne, and Ar). As can be seen from the table data, the relation $H/R\approx 0.2$ holds only for the ground states of the nanobubble.

The dependence of the ratio $H/R$ on temperature $T$ for the ground states of the nanobubble is shown in Fig.~\ref{fg10}. As can be seen from the figure, the value of $H/R$ increases monotonically with temperature, while the spread of values decreases: $H/R$ varies within the intervals [0.151,~0.206] at $T=0$; [0.174, 0.212] at $T=300$; [0.204, 0.220] at $T=600$; and [0.212, 0.222] at $T=900$~K. These estimates are in good agreement with the universal value $H/R=0.204$ obtained in \cite{Jain2017} for a graphene nanobubble containing an ideal gas.
Note that the study \cite{Korneva2023} reported an estimate of $H/R\approx 0.13$ for graphene nanobubbles with argon atoms at $T=2$~K, indicating that the simulation used not the ground stationary state of the nanobubble, but rather a state with a smaller number of layers.

Note that the shape of the nanobubble's ground state depends on the interaction energy $\epsilon_0$ between the graphene sheet and the substrate. 
The temperature dependence of the $H/R$ ratio for three characteristic values of the adhesion energy $\epsilon_0=0.037$, $0.074$, $0.148$~eV is shown in Fig.~\ref{fg11}.
As can be seen from the figure, an increase in the adhesion energy leads to a monotonic increase in the $H/R$ ratio. 
For example, at a temperature of $T=300$~K, the $H/R$ ratio can increase from $0.163$ for weak adhesion to $0.222$ for strong adhesion; at $T=600$~K --- from $0.177$ to $0.238$; and at $T=900$~K --- from $0.182$ to $0.247$.
This monotonic dependence of the $H/R$ ratio on $\epsilon_0$ allows for experimental estimates of the adhesion between the graphene sheet and the substrate \cite{Wang2016}.

\section{Conclusion \label{sec5}}
Using the example of inert gas atoms (He, Ne, Ar, Kr, Xe), it has been shown that graphene nanobubbles
on flat substrates exhibit multistability.
These bubbles can adopt multiple stable stationary states, each defined by a distinct number of layers, $l$, within an internal cluster of atoms.
Within the nanobubble, each atom is confined to one of these layers.
The layers themselves are circular, flat, and concentrically stacked, forming an $l$-stepped pyramid with a flat top.
While atoms within each layer occupy a single plane, the edge atoms are displaced downward, effectively smoothing the stepped profile of the pyramid.
This pyramidal cluster is encapsulated by a locally stretched graphene sheet.
The sheet's valence bonds elongate only directly above the group of confined atoms; outside this coverage zone, the membrane remains undeformed and tightly adhered to the substrate.
The most significant elongation, reaching up to one percent, occurs directly above the layer edges.

The maximum number of possible layers $l_m$ increases monotonically with the total number of encapsulated atoms, $N_g$.
For argon atoms, only single-layer stationary states ($l_m=1$) are possible for $N_g< 145$;
for $145<N_g<550$, both single-layer and double-layer states exist ($l_m=2$);
for $550<N_g<1600$, triple-layer states are added ($l_m=3$);
for $1600<N_g<2500$, four-layer states appear ($l_m=4$);
and five-layer states ($l_m=5$) become accessible for $2500<N_g<4000$.
The graphene membrane, through van der Waals interaction with the substrate, exerts pressure on the encapsulated atomic cluster, confining it against the substrate with pressures on the order of $P\sim 1$ GPa.
This pressure is most pronounced for single-layer atomic packings and systematically decreases as the number of layers increases.

Thermal fluctuation modeling reveals that among all $l$-layer states of a graphene nanobubble, one "ground"\ state always exist.
This state smoothly transitions into a layerless liquid configuration as the temperature increases, whereas the remaining $l$-layer states undergo a transition into this ground state at a distinct temperature $T_l$.
For a cluster of $N_g=4000$ encapsulated atoms, the ground state corresponds to the four-layer packing ($l=4$).
The thermal stability of $l$-layer states varies significantly with the atomic species.
For helium atoms, the $l$-layer states persist up to temperatures $T_l = 290$, $72$, $47$, and $23$~K for $l = 1$, $2$, $3$, and $5$, respectively.
For neon, the corresponding stability limits are 390, 120, 90, and 60~K;
for argon---630, 270, 180, and 120~K;
for krypton---720, 300, 240, and 180~K; and for xenon---780, 390, 330, and 240~K.

The coexistence of multiple stable states with distinct layer numbers at low temperatures demonstrates that universal scaling does not apply to these graphene nanobubbles.
Specifically, the height-to-radius ratio $H/R$ is not constant and can vary across the range from 0 to 0.28.
The commonly reported universal relation $H/R\approx 0.2$ holds only for the ground states of the system.
\\ \\
{\bf Acknowledgements}\\

Computational facilities were provided by the Joint Supercomputer center (JSCC) of the National Research Center "Kurchatov Institute".
The research was funded by the Russian Science Foundation (RSF) (project No. 25-73-20038).
\\ \\
\section*{Data Availability}

All data are openly available \cite{data}


\begin{references}
\bibitem{Novoselov2004}%
K. S. Novoselov, A. K. Geim, S. V. Morozov, D. Jiang, Y. Zhang, S. V. Dubonos, I. V. Grigorieva, A. A. Firsov.
Electric Field Effect in Atomically Thin Carbon Films.
Science {\bf 306}(5696), 666-669 (2004).
https://doi.org/10.1126/science.1102896
\bibitem{Geim2007}
A. K. Geim, K. S. Novoselov.
The rise of graphene.
Nat. Mater. {\bf 6}(3), 183-191 (2007).
https://doi.org/10.1038/nmat1849
\bibitem{Soldano2010}
C. Soldano, A. Mahmood, E. Dujardin.
Production, Properties and Potential of Graphene.
Carbon {\bf 48}(8), 2127-2150 (2010).
https://doi.org/10.1016/j.carbon.2010.01.058
\bibitem{Baimova2014a}
J. A. Baimova, B. Liu, S. V. Dmitriev, K. Zhou.
Mechanical properties and structures of bulk nanomaterials based on carbon nanopolymorphs.
Phys. Status Solidi RRL {\bf 8}(4), 336-340 (2014).
https://doi.org/10.1002/pssr.201409063
\bibitem{Baimova2014b}
J. A. Baimova, E. A. Korznikova, S. V. Dmitriev, B. Liu and  K. Zhou.
Review on crumpled graphene: Unique mechanical properties.
Rev. Adv. Mater. Sci. {\bf 39}, 69-83 (2014).
https://www.ipme.ru/e-journals/RAMS/no\_13914/11\_13914\_baimova.pdf
\bibitem{Geim2009}
A. K. Geim.
Graphene: status and prospects.
Science {\bf 324}(5934), 1530-1534 (2009).
https://doi.org/10.1126/science.1158877
\bibitem{Lee2008}
C. Lee, X. Wei, J.W. Kysar, J. Hone.
Measurement of the elastic properties and intrinsic strength of monolayer graphene.
Science {\bf 321}(5887), 385-388 (2008).
https://doi.org/10.1126/science.1157996
\bibitem{Baladin2008}
A. A. Balandin, S. Ghosh, W. Bao, I. Calizo, D. Teweldebrhan, F. Miao, C. N. Lau.
Superior thermal conductivity of single-layer graphene.
Nano Lett. {\bf }8(3), 902-907 (2008).
https://doi.org/10.1021/nl0731872
\bibitem{Liu2015}
Y. Liu, C. Hu, J. Huang, B. G. Sumpter, R. Qiao.
Tuning interfacial thermal conductance of graphene embedded in soft materials by vacancy defects.
J. Chem. Phys. {\bf 142}(24), 244703 (2015).
https://doi.org/10.1063/1.4922775
\bibitem{Geim2013}
A. K. Geim and I. V. Grigorieva.
Van der Waals heterostructures.
Nature {\bf 499}, 419-425 (2013).
https://doi.org/10.1038/nature12385
\bibitem{Wang2013}
L. Wang, I. Meric,  P.Y. Huang,  Q. Gao, Y. Gao, H. Tran, T. Taniguchi, K. Watanabe,
L.M. Campos, D. A. Muller, J. Guo, P. Kim, J. Hone, K.L. Shepard, and C.R. Dean.
One-dimensional electrical contact to a two-dimensional material.
Science {\bf 342}, 614-617 (2013).
https://doi.org/10.1126/science.1244358
\bibitem{Xiang2020}
R. Xiang, T. Inoue, Y. Zheng, A. Kumamoto, Y. Qian, Y. Sato, M. Liu, D. Tang, D. Gokhale, J. Guo,
K. Hisama, S. Yotsumoto, T. Ogamoto, H. Arai, Y. Kobayashi, H. Zhang, B. Hou, A. Anisimov, M. Maruyama,
Y. Miyata, S. Okada, S. Chiashi, Y. Li, J. Kong, E.I. Kauppinen, Y. Ikuhara, K. Suenaga, S. Maruyama.
One-dimensional van der Waals heterostructures.
Science {\bf 367}, 537-542 (2020).
https://doi.org/10.1126/science.aaz2570
\bibitem{Zhang2020}
Y. Zhang, C. Hu, B. Lyu, H. Li, Z. Ying, L. Wang, A. Deng, X. Luo, Q. Gao, J. Chen, J. Du, P. Shen,
K. Watanabe, T. Taniguchi, J.-H. Kang, F. Wang, Y. Zhang, and Z. Shi.
Tunable cherenkov radiation of phonon polaritons in silver nanowire/hexagonal boron nitride heterostructures.
Nano Lett.  {\bf 20}, 2770-2777 (2020).
https://doi.org/10.1021/acs.nanolett.0c00419
\bibitem{Vasu2016}
K.S. Vasu, E. Prestat, J. Abraham, J. Dix, R.J. Kashtiban, J. Beheshtian, J. Sloan, P. Carbone,
M. Neek-Amal, S.J. Haigh, A.K. Geim, and R.R. Nair.
Van der Waals pressure and its effect on trapped interlayer molecules.
Nat. Commun. {\bf 7}, 12168 (2016).
https://doi.org/10.1038/ncomms12168
\bibitem{Khestanova2016}
E. Khestanova, F. Guinea, L. Fumagalli, A.K. Geim, and I.V. Grigorieva.
Universal shape and pressure inside bubbles appearing in van der Waals heterostructures.
Nat. Commun. {\bf 7}, 12587 (2016).
https://doi.org/10.1038/ncomms12587
\bibitem{Hu2023}
C. Hu, J. Chen, X. Zhou, Y. Xie, X. Huang, Z. Wu, S. Ma, Z. Zhang, K. Xu, N. Wan, Y. Zhang,
Q. Liang, and  Z. Shi.
Collapse of carbon nanotubes due to local high-pressure from van der Waals encapsulation.
Nat. Commun. {\bf 15}, 3486 (2024).
https://doi.org/10.1038/s41467-024-47903-3
\bibitem{Zhang2017}
L. Zhang, Y. Wang, J. Lv, and Y. Ma.
Materials Discovery at High Pressures.
Nature Reviews Materials, {\bf 2}, 17005 (2017).
https://doi.org/10.1038/natrevmats.2017.5
\bibitem{Zheng2012}
J. Zheng, X. Liu, P. Xu, P. Liu, Y. Zhao, and J. Yang.
Development of high pressure gaseous hydrogen storage technologies.
International Journal of Hydrogen Energy, {\bf 37}(1), 1048-1057 (2012).
https://doi.org/10.1016/j.ijhydene.2011.02.125
\bibitem{Slepchenkov2018}
M. M. Slepchenkov, P. V. Barkov and O. E. Glukhova.
High-Density Hydrogen Storage in a 2D-Matrix from Graphene Nanoblisters:
A Prospective Nanomaterial for Environmentally Friendly Technologies.
Crystals, {\bf 8}(4), 161 (2018).
https://doi.org/10.3390/cryst8040161
\bibitem{Apkadirova2022}
N.G. Apkadirova, K.A. Krylova, J.A. Baimova.
Effect of external pressure on the hydrogen storage capacity of a graphene flake: molecular dynamics.
Lett. Mater., {\bf 12}(4s), 445-450 (2022).
https://doi.org/10.22226/2410-3535-2022-4-445-450
\bibitem{Wang2016}
J. Wang, D. C. Sorescu, S. Jeon, A. Belianinov, S. V. Kalinin, A. P. Baddorf, and P. Maksymovych.
Atomic intercalation to measure adhesion of graphene on graphite.
Nat Commun {\bf 7}, 13263 (2016). 
https://doi.org/10.1038/ncomms13263
\bibitem{He2023}
Y. He, Y. Dong, Y. Zhang, Y. Li, and H. Li.
Graphene Nano-Blister in Graphite for Future Cathode in Dual-Ion Batteries: Fundamentals, Advances, and Prospects.
Adv. Sci. {\bf 10}, 2207426 (2023).
https://doi.org/10.1002/advs.202207426
\bibitem{Langle2024}
M. Langle, K. Mizohata, C. Mangler, A. Trentino, K. Mustonen, E. H. Ahlgren, and J. Kotakoski.
Two-dimensional few-atom noble gas clusters in a graphene sandwich.
Nat. Mater. {\bf 23}, 762-767 (2024). 
https://doi.org/10.1038/s41563-023-01780-1
\bibitem{Sanchez2018}
D. A. Sanchez, Z. Dai, P. Wang, A. Cantu-Chavez, C. J. Brennan, R. Huang and N. Lu.
Mechanics of spontaneously formed nanoblisters trapped by transferred 2D crystals.
Proceedings of the National Academy of Sciences, {\bf 115}(31), 7884-7889 (2018).
https://doi.org/10.1073/pnas.1801551115
\bibitem{Blundo2020}
E. Blundo, C. Di Giorgio, G. Pettinari, T. Yildirim, M. Felici, Y. Lu, F. Bobba and A. Polimeni.
Engineered Creation of Periodic Giant, Nonuniform Strains in MoS2 Monolayers.
Advanced Materials Interfaces, {\bf 7}(17), 2000621 (2020).
https://doi.org/10.1002/admi.202000621
\bibitem{Zamborlini2015}
G. Zamborlini, M. Imam, L. L. Patera, T. O. Mentes, N. Stojic, C. Africh, A. Sala, N. Binggeli,
G. Comelli and  A. Locatelli,
Nanobubbles at GPa pressure under graphene.
Nano  Lett., {\bf 15}, 6162-6169 (2015).
https://doi.org/10.1021/acs.nanolett.5b02475
\bibitem{Villarreal2021}
R. Villarreal, P.-C. Lin, F. Faraji, N. Hassani, H. Bana, Z. Zarkua, M. N. Nair, H.-C. Tsai,
M. Auge, F. Junge, H. C. Hofsaess, S. De Gendt, S. De Feyter, S. Brems, E. H. Ahlgren, E. C. Neyts,
L. Covaci, F. M. Peeters, M. Neek-Amal, and  L. M. C. Pereira.
Breakdown of Universal Scaling for Nanometer-Sized Bubbles in Graphene.
Nano Lett., {\bf 21}, 8103-8110 (2021).
https://doi.org/10.1021/acs.nanolett.1c02470
\bibitem{Zahra2020}
K. M. Zahra, C. Byrne, A. Alieva, C. Casiraghi and A. S. Walton.
Intercalation, decomposition, entrapment - a new route to graphene nanobubbles.
Phys. Chem. Chem. Phys. {\bf 22}(14), 7606-7615 (2020).
https://doi.org/10.1039/d0cp00592d.
\bibitem{Yue2012}
K. Yue, W. Gao, R. Huang and K. M. Liechti.
Analytical methods for the mechanics of graphene bubbles
J. Appl. Phys. {\bf 112}, 083512 (2012).
https://doi.org/10.1063/1.4759146
\bibitem{Wang2013a}
P. Wang, W. Gao, Z. Cao, K. M. Liechti and R. Huang.
Numerical Analysis of Circular Graphene Bubbles.
Journal of  Applied  Mechanics, {\bf 80}, 040905 (2013).
https://doi.org/10.1115/1.4024169
\bibitem{Dai2018}
Z. Dai, Y. Hou, D. A. Sanchez, G. Wang, C. J. Brennan, Z. Zhang, L. Liu and N. Lu.
Interface-Governed Deformation of Nanobubbles and Nanotents Formed by Two-Dimensional Materials.
Phys. Rev. Lett., {\bf 121}, 266101 (2018).
https://doi.org/10.1103/PhysRevLett.121.266101
\bibitem{Zhilyaev2019}
P. Zhilyaev, E. Iakovlev and I. Akhatov.
Liquid-gas phase transition of Ar inside graphene nanobubbles on the graphite.
Nanotechnology,  {\bf 30}(21),  215701 (2019).
https://doi.org/10.1088/1361-6528/ab061f
\bibitem{Iakovlev2019}
E. Iakovlev, P. Zhilyaev and  I. Akhatov.
Modeling of the phase transition inside graphene nanobubbles filled with ethane.
Phys. Chem. Chem. Phys., {\bf 21}, 18099-18104 (2019).
https://doi.org/10.1039/c9cp03461g
\bibitem{Aslymov2020}
T. Aslyamov, E. Iakovlev, I. S. Akhatov and P. Zhilyaev.
Model of graphene nanobubble: Combining classical density functional and elasticity theories.
J. Chem. Phys. {\bf 152}, 054705 (2020).
https://doi.org/10.1063/1.5138687
\bibitem{Qu2023}
Z.-X. Qu, B.-S. Wang, and J.-W. Jiang.
The Gas in Graphene Bubbles: An Improved van der Waals Equation Description.
J. Phys. Chem. C, {\bf 127}, 9205-9212 (2023).
https://doi.org/10.1021/acs.jpcc.2c08173
\bibitem{Lyublinskaya2020}
A. Lyublinskaya, S. Babkin and I. Burmistrov.
Effect of anomalous elasticity on bubbles in van der Waals heterostructures.
Physical  Review E, {\bf 101}, 033005 (2020)
https://doi.org/10.1103/PhysRevE.101.033005
\bibitem{Iakovlev2017}
E. Iakovlev, P. Zhilyaev and I. Akhatov.
Atomistic study of the solid state inside graphene nanobubbles.
Scientific reports, {\bf 7}, 17906 (2017).
https://doi.org/10.1038/s41598-017-18226-9
\bibitem{Ghorbanfekr2017}
H. Ghorbanfekr-Kalashami, K. Vasu, R. R. Nair,  F. M. Peeters and M. Neek-Amal.
Dependence of the shape of graphene nanobubbles on trapped substance.
Nature  communications,  {\bf 8}, 1-11 (2017).
https://doi.org/10.1038/ncomms15844
\bibitem{Jain2017}
S. K. Jain, V. Juricic,  and  G. T. Barkema.
Probing the Shape of a Graphene Nanobubble.
Phys. Chem. Chem. Phys., {\bf 19}, 7465-7470 (2017).
https://doi.org/10.1039/C6CP08535K
\bibitem{Korneva2023}
M. Korneva  and  P. Zhilyaev.
Solid-liquid phase transition inside Van der Waals nanobubbles: tomistic perspective.
Phys. Chem. Chem. Phys., {\bf 25}, 18788-18796 (2023).
https://doi.org/10.1039/D3CP01285A
\bibitem{Faraji2022}
F. Faraji, M. Neek-Amal, E. C. Neyts, and F. M. Peeters.
Indentation of graphene nano-bubbles.
Nanoscale, {\bf 14}, 5876-5883 (2022).
https://doi.org/10.1039/D2NR01207C
\bibitem{Qu2023a}
Z.-X. Qu and J.-W. Jiang.
Nanobubble-induced significant reduction of the interfacial thermal conductance for few-layer graphene
Phys. Chem. Chem. Phys., {\bf 25}, 28651 (2023).
https://doi.org/10.1039/d3cp04085b
\bibitem{Qu2023b}
Z.-X. Qu, C.-X. Cui, J.-W. Jiang.
Bubble-Induced Strong Thermal Contraction for  Graphene.
ASME Journal of Heat and Mass Transfer, {\bf 145}(12), 1-13 (2023).
https://doi.org/10.1115/1.4063230
\bibitem{Savin2010}
A. V. Savin, Y. Kivshar, and B. Hu.
Suppression of thermal conductivity in graphene nanoribbons with rough edges.
Phys. Rev. B, {\bf 82}, 195422 (2010).
https://doi.org/10.1103/PhysRevB.82.195422
\bibitem{Savin2008}
A. V. Savin and Yu. S. Kivshar.
Discrete breathers in carbon nanotubes.
EPL {\bf 82}(6), 66002 (2008).
https://doi.org/10.1209/0295-5075/82/66002
\bibitem{Aitken2010}
Z. H. Aitken, R. Huang.
Effects of mismatch strain and substrate surface corrugation on morphology of supported monolayer graphene.
J. Appl. Phys. {\bf 107}, 123531 (2010).
https://doi.org/10.1063/1.3437642
\bibitem{Zhang2013}
K. Zhang and M. Arroyo.
Adhesion and friction control localized folding in supported graphene.
J. Appl. Phys. {\bf 113}, 193501 (2013).
https://doi.org/10.1063/1.3437642
\bibitem{Zhang2014}
K. Zhang, M. Arroyo.
Understanding and strain-engineering wrinkle networks in supported graphene through simulations.
Journal of the Mechanics and Physics of Solids {\bf 72}, 61-74 (2014).
https://doi.org/10.1016/j.jmps.2014.07.012
\bibitem{Koenig2011}
S. P. Koenig, N. G. Boddeti, M. L. Dunn, J. S. Bunch.
Ultrastrong adhesion of graphene membranes.
Nature Nanotech {\bf 6}, 543-546 (2011).
https://doi.org/10.1038/nnano.2011.123
\bibitem{Rappe1992}
A. K. Rappe, C. J. Casewit, K. S. Colwell, W. A. Goddard III, W. M. Skiff.
UFF, a Full Periodic Table Force Field for Molecular Mechanics and Molecular Dynamics Simulations.
J. Am. Chem. Soc. {\bf 114}(25), 10024-10035 (1992).
https://doi.org/10.1021/ja00051a040
\bibitem{Vogt2014}
J. Vogt and S. Alvarez.
van  der  Waals  Radii  of  Noble  Gases.
Inorg. Chem. {\bf 53}(17), 9260-9266 (2014).
https://doi.org/10.1021/ic501364h
\bibitem{Lindsay2010}
L. Lindsay and D. A. Broido.
Optimized  tersoff and brenner empirical potential parameters for lattice dynamics and phonon thermal transport in carbon nanotubes and graphene.
Phys. Rev. B {\bf 81}, 205441  (2010).
https://doi.org/10.1103/PhysRevB.81.205441
\bibitem {Stuart2000}
S. J. Stuart, A. B. Tutein, J. A. Harrison.
A reactive potential for hydrocarbons with intermolecular interactions.
J. Chem. Phys. {\bf 112} (14), 6472-6486 (2000).
https://doi.org/10.1063/1.481208
\bibitem{Brener2002}
D. W. Brenner, O. A. Shenderova, J. A. Harrison, S. J. Stuart, B. Ni, S. B. Sinnott.
A second-generation reactive empirical bond order (rebo) potential energy expression for hydrocarbons,
J. Phys.: Condens. Matter. {\bf 14} (4), 783-802 (2002).
https://doi.org/10.1088/0953-8984/14/4/312
\bibitem{Srinivasan2015}
S. G. Srinivasan, A. C. T. van Duin, P. Ganesh.
Development of a reaxff potential for carbon condensed phases and its application to the thermal fragmentation of large fullerene.
J. Phys. Chem. A {\bf 119}, 571-580 (2015).
https://doi.org/10.1021/jp510274e
\bibitem{Lebedeva2019}
I. V. Lebedeva, A. S. Minkin, A. M. Popov, A. A. Knizhnik.
Elastic constants of graphene: Comparison of empirical potentials and dft calculations.
Phys. E {\bf 108}, 326-338 (2019).
https://doi.org/10.1016/j.physe.2018.11.025
\bibitem{Fletcher1964}
R. Fletcher and C. Reeves.
Function Minimization by Conjugate Gradients.
Computer Journal {\bf 7}(2), 149-154 (1964).
https://doi.org/10.1093/comjnl/7.2.149
\bibitem{Shanno1976}
D. F. Shanno, K. H. Phua.
Algorithm 500: Minimization of Unconstrained Multivariate Functions [E4].
ACM Transactions on Mathematical Software {\bf 2}(1), 87-94 (1976).
https://doi.org/10.1145/355666.355673
\bibitem{Zabel2011}
J. Zabel, R. R. Nair, A. Ott, T. Georgiou, A. K. Geim, K. S. Novoselov, and C. Casiraghi.
Raman Spectroscopy of Graphene and Bilayer under Biaxial Strain: Bubbles and Balloons.
Nano Lett. {\bf 12}(2), 617-621 (2012).
https://doi.org/10.1021/nl203359n
\bibitem{Georgiou2011}
T. Georgiou, L. Britnell, P. Blake, R. V. Gorbachev, A. Gholinia, A. K. Geim, C. Casiraghi, and K. S. Novoselov.
Graphene bubbles with controllable curvature.
Appl. Phys. Lett. {\bf 99}, 093103 (2011).
http://doi.org/10.1063/1.3631632
\bibitem{Savin2019}
A. V. Savin and O. I. Savina.
Bistability of Multiwalled Carbon Nanotubes Arranged on Plane Substrates. 
Physics of the Solid State, {\bf 61}(11), 2241-2248 (2019).
https://doi.org/10.1134/S1063783419110295
\bibitem{Lahiri2011}
J. Lahiri, T. S. Miller, A. J. Ross, L. Adamska, I. I. Oleynik, M. Batzill. 
Graphene growth and stability at nickel surfaces.
New J. Phys. {\bf 13}, 025001 (2011)
https://doi.org/10.1088/1367-2630/13/2/025001
\bibitem{Verlet1967}
L. Verlet.
Computer "Experiments" on Classical Fluids. I. Thermodynamical Properties of Lennard-Jones Molecules.
Phys. Rev. {\bf 159}, 98 (1967).
https://doi.org/10.1103/PhysRev.159.98
\bibitem{data}
A. V. Savin, 
Supplementary Material for the article "Multistability of graphene nanobubbles". 
Zenodo (2026). https://zenodo.org/records/18900382 
\end{references}
\end{document}